\documentclass{aastex631}
\usepackage{graphicx}
\usepackage{natbib}
\usepackage{multirow}
\usepackage{amsmath}
\usepackage{amssymb}
\usepackage{textcomp, gensymb}
\usepackage{rotating}
\usepackage{longtable}

\newcommand{\kms}{km~s$^{-1}$}
\newcommand{\am}{NH$_{3}$}

\newcommand{\cm}{cm$^{-3}$}
\newcommand{\htwo}{H$_2$}

\begin{document}

\title{Widespread Hot Ammonia in the Central Kiloparsec of the Milky Way}

\author{T.M. Candelaria}
\affiliation{New Mexico Institute of Mining and Technology, 801 Leroy Socorro, NM 87801}

\email{tierra.candelaria@student.nmt.edu}

\author{E.A.C. Mills}
\affiliation{Department of Physics and Astronomy, University of Kansas, 1251 Wescoe Hall Dr., Lawrence, KS 66045, USA}

\author{D.S. Meier}
\affiliation{New Mexico Institute of Mining and Technology, 801 Leroy Socorro, NM 87801}

\author{J. Ott}
\affiliation{National Radio Astronomy Observatory, e1003 Lopezville Rd Socorro, NM 87801}

\author{N. Butterfield}
\affiliation{Department of Physics, Villanova University, 800 E. Lancaster Ave., Villanova, PA 19085, USA}

\begin{abstract}

The inner 300-500 pc of the Milky Way has some of the most extreme gas conditions in our Galaxy. Physical properties of the Central Molecular Zone (CMZ), including temperature, density, thermal pressure, and turbulent pressure, are key factors for characterizing gas energetics, kinematics, and evolution. The molecular gas in this region is more than an order of magnitude hotter than gas in the Galactic disk, but the mechanism responsible for heating the gas remains uncertain. We characterize the temperature for 16 regions, extending out to a projected radius of $\sim$450 pc. We observe \am\, J,K=(1,1)-(6,6) inversion transitions from SWAG (Survey of Water and Ammonia in the Galactic Center) using the Australia Telescope Compact Array (ATCA), and ammonia lines (J,K) = (8,8)-(14,14) using the 100\,m Green Bank Telescope. Using these two samples we create full Boltzmann plots for every source and fit two rotational temperature components to the data. For the cool component we detect rotational temperatures ranging from 20-80\,K, and for the hot component we detect temperature ranging from 210-580\,K. With this sample of 16 regions, we identify some of the most extreme molecular gas temperatures detected in the Galactic center thus far. We do not find a correlation between gas temperature and Galactocentric radius, and we confirm that these high temperatures are not exclusively associated with actively star-forming clouds. We also investigate temperature and line widths and find (1) no correlation between temperature and line width and (2) the lines are non-thermally broadened indicating that non-thermal motions are dominant over thermal.

\end{abstract}



\section{Introduction}

Sitting at a distance of 8.1 kpc from the Sun \citep{GRAVITY18,Gravity19}, the Galactic center is one of the most extreme environments in our Galaxy. This innermost region of the Galaxy, also known as the Central Molecular Zone (CMZ), only spans 300-500 pc in diameter, but contains $\sim$5\% of all the molecular gas in the Galaxy \citep{Nakanishi06}. This gas is primarily located in giant molecular clouds with much higher typical densities \citep[$n=10^3-10^5$\, \cm;][] {Guesten83,Bally88,Tsuboi99,RF01} and kinetic temperatures \citep[$T=70-300$ K;][]{Guesten83,Guesten85,Mauers86,Huttem93b,Ginsburg16,Mills13a,Mills18a} than found in other regions of the Galaxy. Looking outside our galaxy, these extreme conditions are more commonly found in many high-redshift galaxies \citep{Tacconi08,FS09,Tacconi10,Genzel10,Swinbank11}. Although, it is not clear whether the physical mechanisms responsible for these gas conditions in high redshift galaxies are also the mechanisms responsible for the gas conditions in the CMZ. 

Many possible heating mechanisms, including UV heating through photodissociation regions \citep{RF04}, gravitational heating \citep{GL78}, X-rays \citep{Maloney96}, cosmic rays \citep{GL78,Ao13}, turbulence \citep{Wilson82,Ao13,Ginsburg16,Goic13,Immer16}, and large-scale instabilities \citep{Kruijssen14} have been suggested to contribute to the heating of the  molecular gas in the CMZ. Identifying the mechanism(s) that are primarily responsible requires a better understanding of the temperature structure of this gas, particularly of the hottest molecular gas. 

\begin{figure*}[tbh]
    \centering
    \includegraphics[width=\textwidth]{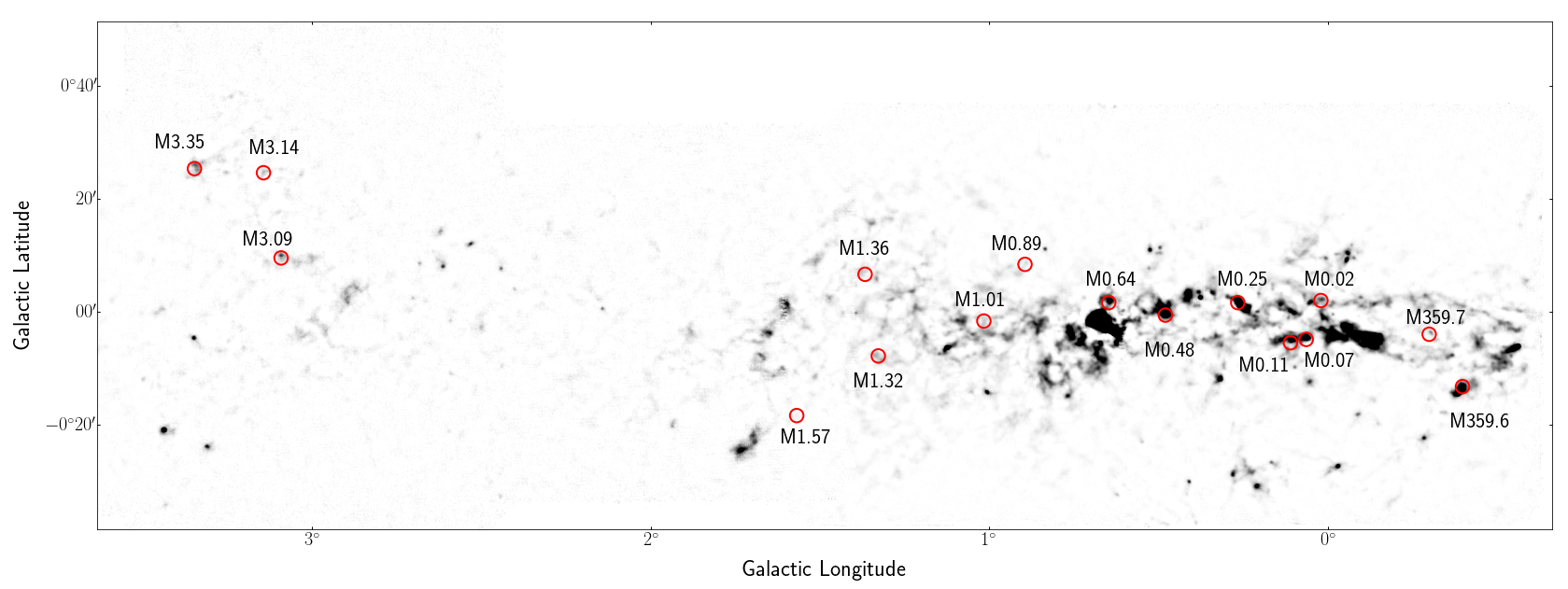}
    \caption{Locations of new measurements of highly-excited ammonia are shown on top of a 1.1 mm map of dust emission from \citealt{Bally10}. The central supermassive black hole Sgr A* is located at Galactic coordinates $[l,b]$ = [359$\degr$.944, -00$\degr$.046]. The points at $l\sim3\degr$ are located in Bania's clump 2, at a projected Galactocentric radius of $\sim$425 pc.}
    \label{finder}
\end{figure*}

The majority of the measurements of high-temperature molecular gas in the CMZ have been made using the ammonia molecule (\am). Ammonia is an easily-thermalized molecule that has a large number of strong, temperature sensitive metastable lines (J=K) that can be efficiently observed via inversion lines at radio frequencies. These properties make \am\, a highly effective temperature probe that is sensitive to a wide range of excitation conditions in diverse environments \citep{Ho83,Walmsley83,Danby88,Huttem93b}. But while the \am\, thermometer has been shown to be robust for temperatures $<$300\,K \citep{Danby88,Maret09,Bouhafs17}, it has not been rigorously explored at high temperatures.

Previous studies using \am\, have found evidence of multiple temperature components present in CMZ gas. A ``cool" gas component primarily seen in the (1,1) and (2,2) lines has temperatures that range from $\sim$25-60\,K \citep{Mauers86,Huttem93b,Huttem95,Ott14,Lu17,Krieger17}, while a ``warm" component traced by the (3,3) to (5,5) or (6,6) lines has temperatures around 100-150\,K \citep{Gusten85,Huttem95,Lu17,Krieger17}. In several clouds even higher-excitation lines up to (18,18) have been observed. These lines have been used to characterize a ``hot" gas component that has temperatures ranging from 200-600 K \citep{Mauers86,Huttem95,Wilson06,Mills13a}. 

\am\, is not the only molecule that appears to trace hot molecular gas in the CMZ. Warm gas temperatures ($\sim$60-120\,K) measured with \am\, have been compared with those measured from CH$_3$CN and CH$_3$CCH in a larger sample of CMZ clouds, and found to be consistent \citep{Guesten85}. Hot gas temperatures have also been measured in CMZ clouds with other molecules including H$_2$ \citep[T$>$ 600 K;][]{RF01,Mills17b} and CO[T$\sim$300-1000 K;] \citep{Goicoechea13,Etxaluze13}. At present however, Sgr B2 is the only cloud where measurements of the hot molecular gas exist using multiple tracers. Ultimately, while \am\, is generally regarded as a reliable temperature probe, its behavior at the high temperature end and in extreme environments like the CMZ needs to be further explored.

In this paper, we expand on the work of \cite{Mills13a} by observing a larger sample of high excitation \am\, transition toward 16 additional clouds in the CMZ in the (8,8) through (14,14) lines to measure rotational temperatures in these regions. Additionally we have added lower excitation transitions, (1,1)-(6,6), in these 16 sources to determine the rotational temperature from the lower J=K transitions. In Section \ref{obs_cal} we present our observations for three data sets and describe the data calibration process. Section \ref{spectra} describes our detections and Section \ref{temperatures} derives temperatures and the hot gas fraction. Lastly, in Section \ref{discussion} we assess various possibilities for the existence of substantial amounts of highly-excited \am\, emission and address possible heating mechanisms consistent with our results.

\section{Observations and Calibration}
\label{obs_cal}

This paper consists of four data sets, two taken with the 100\,m Green Bank Telescope (GBT)\footnote{The Green Bank Observatory is a facility of the National Science Foundation operated under cooperative agreement by Associated Universities, Inc.}, one taken with the Australia Telescope Compact Array (ATCA)\footnote{The Australia Telescope Compact Array (ATCA) is operated by CSIRO at the Paul Wild Observatory and is a part of the Australia Telescope National Facility network of radio telescopes}, and the last taken with the Mopra Telescope\footnote{The Mopra telescope, located in the Warrumbungle Mountains, is operated by CSIRO's Astronomy and Space Science division}. The GBT data sets were taken in 2014 and 2020, referred to as GBT 2014 and GBT 2020 for the remainder of the paper. The third set of data, taken with the ATCA, are provided by SWAG: ``Survey of Water and Ammonia in the Galactic Center'' \citep{Krieger17}. The remaining data is taken from the H$_2$O Southern Galactic Plane Survey (HOPS) (\citep{HOPS1,HOPS2,HOPS3}) and will be referred to for the remainder of the paper as HOPS.

\begin{table}[t]
\centering
\caption{Observed Transitions of \am} 
\begin{tabular}{cccc}
\hline\hline
\bf{Transition} & \bf{Rest } & \bf{Upper } & \bf{Beam Size}\\
 & \bf{Frequency} & \bf{State} & \\
  & & \bf{Energy}$^{\mathrm a}$ & \\
{\bf (J,K)} & {\bf (GHz)} & {\bf (K)} & {\bf ($''$)}\\
\hline
(1,1)$^{\mathrm b}$ & {23.6945} & {23.3} & {26.2 $\times$ 17.8}  \\
(2,2)$^{\mathrm b}$ & 23.7226 & 64.4 & 26.2 $\times$ 17.8 \\
(3,3)$^{\mathrm b}$ & 23.8701 & 123.5 & 26.0 $\times$ 17.7 \\
(4,4)$^{\mathrm b}$ & 24.1394 & 200.5 & 25.7 $\times$ 17.5 \\
(5,5)$^{\mathrm b}$ & 24.5329 & 295.4 & 25.3 $\times$ 17.2 \\
(6,6)$^{\mathrm b}$ & 25.0560 & 408.1 & 24.8 $\times$ 16.9 \\
\hline
(8,8)$^{\mathrm c}$ & 26.5189 & 686.8 & 28.5 \\
(9,9)$^{\mathrm c}$ & 27.4779 & 852.8 & 27.5 \\
(10,10)$^{\mathrm c}$ & 28.6046 & 1036.4 & 26.4 \\
(11,11)$^{\mathrm c}$ & 29.9145 & 1237.6 & 25.2 \\
(12,12)$^{\mathrm c}$ & 31.4249 & 1456.4 & 24.0 \\
(13,13)$^{\mathrm c}$ & 33.1568 & 1692.7 & 22.8 \\
(14,14)$^{\mathrm c}$ & 35.1343 & 1944.6 & 21.5 \\
\hline
\end{tabular}
\label{table:ammonia}

$^{\mathrm a}$Values taken from the JPL Submillimeter, Millimeter, and Microwave Spectral Line Catalog \citep{Pickett98}\\
$^{\mathrm b}$Interferometric data from the SWAG survey \citep{Krieger17}
$^{\mathrm c}$Single-dish GBT data
\end{table}

\begin{table*}[th]
\caption{Observed Sources}
\centering
\begin{tabular}{lcccccc}
\\[0.5ex]
\hline\hline
 {\bf Source} & {\bf Location} & {\bf R.A.} & {\bf Decl.} & {\bf Galactic} & {\bf Galactic} & {\bf Projected }\\ 
 & & {  \bf (J2000)} & {\bf (J2000)} & {\bf Longitude} & {\bf Latitude} & {\bf Galactocentric} \\
 & & & & & & {\bf Distance (pc)} \\

\hline
{\bf M3.35+0.43} & {Bania's Clump 2} & 17$^{\mathrm{h}}$51$^{\mathrm{m}}$44.92$^{\mathrm{s}}$ & -25\degr 51\arcmin 01.42\arcsec &   3\degr.3453 & 0\degr.4256  & {470} \\

{\bf M3.14+0.41} & {Bania's Clump 2} &  17$^{\mathrm{h}}$51$^{\mathrm{m}}$19.88$^{\mathrm{s}}$ & -26\degr 01\arcmin 54.89\arcsec &   3\degr.1413 & 0\degr.4135  & {440} \\

{\bf M3.09+0.16} & {Bania's Clump 2} & 17$^{\mathrm{h}}$52$^{\mathrm{m}}$10.71$^{\mathrm{s}}$ & -26\degr 12\arcmin 18.84\arcsec &   3\degr.0893 & 0\degr.1616  & {430}\\

{\bf M1.57-0.30} & {} & 17$^{\mathrm{h}}$50$^{\mathrm{m}}$28.88$^{\mathrm{s}}$ & -27\degr 45\arcmin 10.22\arcsec &   1\degr.5658 & -0\degr.3043  & {220} \\

{\bf M1.36+0.11}$^{\mathrm{a}}$ & {} & 17$^{\mathrm{h}}$48$^{\mathrm{m}}$23.63$^{\mathrm{s}}$ & -27\degr 42\arcmin 41.16\arcsec &   1\degr.3637 & 0\degr.1132  & {190} \\

{\bf M1.32-0.13}$^{\mathrm a}$ & {} &  17$^{\mathrm{h}}$49$^{\mathrm{m}}$14.25$^{\mathrm{s}}$ & -27\degr 52\arcmin 8.93\arcsec &   1\degr.3247 & -0\degr.128  & {180} \\

{\bf M1.01+0.02} & {} & 17$^{\mathrm{h}}$48$^{\mathrm{m}}$6.54$^{\mathrm{s}}$ & -28\degr 04\arcmin 59.8\arcsec &   1\degr.0128 & -0\degr.0249  & {140} \\

{\bf M0.89+0.14} & {} &   17$^{\mathrm{h}}$47$^{\mathrm{m}}$10.4$^{\mathrm{s}}$ & -28\degr 06\arcmin 1.85\arcsec &   0\degr.8913 & 0\degr.1427  & {120} \\

{\bf M0.64-0.03} & {Sgr B2 Complex} & 17$^{\mathrm{h}}$47$^{\mathrm{m}}$01.77$^{\mathrm{s}}$ & -28\degr 22\arcmin 15.19\arcsec &   0\degr.6437 & 0\degr.0296  & {80} \\

{\bf M0.48-0.01}$^{\mathrm{b}}$ & {Dust Ridge (Cloud E/F)} & 17$^{\mathrm{h}}$46$^{\mathrm{m}}$46.71$^{\mathrm{s}}$ & -28\degr 31\arcmin 58.85\arcsec &   0\degr.4765 & -0\degr.0073  & {60} \\

{\bf M0.25+0.01}$^{\mathrm{a}}$ & {Dust Ridge (Brick)} & 17$^{\mathrm{h}}$46$^{\mathrm{m}}$07.70$^{\mathrm{s}}$ & -28\degr 41\arcmin 47.68\arcsec &   0\degr.2626 & 0\degr.0296  &  {30} \\

{\bf M0.11-0.08} & {} & 17$^{\mathrm{h}}$46$^{\mathrm{m}}$13.30$^{\mathrm{s}}$ & -28\degr 53\arcmin 29.00\arcsec &   0\degr.1068 & -0\degr.0891  & {10} \\

{\bf M0.07-0.08} & {} & 17$^{\mathrm{h}}$46$^{\mathrm{m}}$04.36$^{\mathrm{s}}$ & -28\degr 55\arcmin 33.32\arcsec &   0\degr.0603 & -0\degr.0793  & {5} \\

{\bf M0.02+0.04} & {} & 17$^{\mathrm{h}}$45$^{\mathrm{m}}$31.27$^{\mathrm{s}}$ & -28\degr 54\arcmin 11.1\arcsec &   0\degr.0169 & 0\degr.0357  & {13} \\

{\bf M359.7+0.064} & {} & 17$^{\mathrm{h}}$45$^{\mathrm{m}}$8.83$^{\mathrm{s}}$ & -28\degr 13\arcmin 40.99\arcsec &   359\degr.6970 & -0\degr.064  & {50} \\

{\bf M359.6+0.22}$^{\mathrm{a}}$ & {}  & 17$^{\mathrm{h}}$45$^{\mathrm{m}}$31.01$^{\mathrm{s}}$ & -29\degr 23\arcmin 40.33.58\arcsec &   359\degr.5986 & -0\degr.2186  & {70} \\

\hline
\end{tabular}
\label{table:sources}

$^{\mathrm a}$Sources exhibiting multiple velocity components
$^{\mathrm b}$Sources with resolved hyperfine splitting

\end{table*}

\subsection{GBT 2014 Observations}
Using the Ka-Band receiver (26.0-39.5 GHz) with VEGAS (the Versatile GBT Astronomical Spectrometer) we observed six \am\, metastable inversion lines toward 12 positions in and near the CMZ. The observed transitions include the $(J,K)=$(8,8) through (13,13) transitions of \am. Properties of transitions are given in Table \ref{table:ammonia}.

In this paper we present data for the eight strongest sources in this data set for which at least four transitions could be detected. These sources and their locations are displayed in Table \ref{table:sources}. The VEGAS spectrometer has a simultaneous bandwidth of $\sim$4\,GHz, therefore to cover all of the lines of interest, observations were made in two frequency windows. The first covers a range of 26.2-30.2\,GHz which includes the (8,8)-(11,11) transitions. The second window covers a range of 30.9-33.7\,GHz including the (12,12) and (13,13) transitions. In each window, there are four subbands, each with $\sim$850\,MHz bandwidth and a spectral resolution of 0.4283\,MHz or 0.7\,km/s.

\subsection{GBT 2014 Calibration}
We used the GBTIDL software to calibrate and analyze the spectra, following the method outlined in more detail in \cite{Mills13a}. The observations employed position switching to emission-free positions out of the plane using one of the two single-polarization beams of the Ka-Band receiver. We corrected the antenna temperature for each source for the frequency-dependent opacity at the observed elevation. The atmospheric opacity estimate was taken from a GBT archive of frequency-dependent opacities calculated from the weather conditions at the time of the observations. An approximate amplitude calibration can be performed using the GBT noise diodes. However, for increased accuracy, we also apply an additional amplitude calibration from observations of the standard flux calibrator 3C286. Unfortunately, the flux calibrator data for our observations of the (12,12) and (13,13) lines were not usable, and so for these lines we are limited to the flux accuracy from the noise diodes. The (14,14) line was not observed in this data set. For the remaining lines, we derived a scaling factor using the theoretical aperture efficiency as a function of frequency for the GBT: 

\begin{equation}
    \eta_\mathit{eff,theor} = 2 \times exp (-[0.00922 \cdot \nu (GHz)]^{2})
    \label{theor_eff}
\end{equation}{}

This is compared to the observed aperture efficiency measured from 3C286 using: 

\begin{equation}
    \eta_\mathit{eff,meas} = T_{a} exp(\tau_{atm}/ \text{sin}\,\theta)/S_{\nu}
    \label{meas_eff}
\end{equation}

\noindent where T$_a$ is the antenna temperature, S$_{\nu}$ is the expected flux density, $\theta$ is the elevation at which the flux calibrator was observed, and $\tau_{atm}$ is the atmospheric opacity estimate.

The ratio of the theoretical to measured aperture efficiency yields a frequency dependent scaling factor for the amplitude correction. Observed scaling factors ranged from 0.616 to 1.499. Based on these magnitudes of these corrections, we conservatively assume that the flux scale determined from the noise diode for the (12,12) and (13,13) lines may be uncertain by up to $\pm$50\%. Finally, in order to convert the measured antenna temperatures to main-beam brightness temperatures, we applied a correction for the beam efficiency. This correction for the GBT is 1.32 and is not frequency dependent\footnote{See https://www.gb.nrao.edu/GBT/Performance/}.

\subsection{GBT 2020 Observations}
An additional eight positions were chosen due to their high temperatures in the \am\, (2,2)/(4,4) ratio map obtained from the SWAG data, with a selection criterion that they were at least twice as hot the surrounding gas temperature. These regions are not well studied in the literature, and are associated with various potential sources of potential feedback. We present data from all eight sources with each having detections of at least four lines. The position and associated potential feedback sources are listed in Appendix \ref{notes_on_sources}.

We used the Ka-Band receiver (26.0-39.5 GHz) to observe the $(J,K) = $(8,8) through (14,14) transitions of \am. Table \ref{table:ammonia} gives the properties of the observed transitions including the frequency, upper energy, and beam size. The data were observed using positions switching to emission-free position out of the plane using one of two single-polarization beams of the Ka-Band receiver. The off position was located $\sim$2\arcmin\, from the source and varied for each source. The spectrometer has a simultaneous bandwidth of $\sim$4\,GHz, thus in order to cover the lines of interest we chose two frequency windows. The first window covers the (8,8)-(11,11) transitions corresponding to frequency of 26.2-30.2\,GHz. The second window includes the (12,12)-(14,14) transitions which correspond to frequency 31.2-35.2\,GHz.

\subsection{GBT 2020 Calibration}
The GBT 2020 calibration was performed in a similar manner to the 2014 calibration and \cite{Mills13a}, though without observations of a flux calibrator. We average scans of the same sources observed across different nights and perform a baseline subtraction. We correct for atmospheric opacity using archival weather data and perform an amplitude calibration using the GBT noise diodes. The amplitude calibration is limited to the accuracy achievable with just the noise diode. Lastly, we apply a correction for the beam efficiency in order to convert the measured antenna temperatures to main-beam brightness temperatures as done in the GBT 2014 data. 

\subsection{SWAG Observations}
For the sources located in the inner 3 degrees of the Galaxy (which excludes sources in Bania's Clump 2), we also include SWAG data covering the lower-$J$ (1,1)-(6,6) \am\, transitions. Properties of these \am\, transitions are also given in Table \ref{table:ammonia}. These data were observed over three years (2014-2016) with five 22-m dishes in the most compact configuration of the ATCA \citep{Krieger17}. SWAG mapped the inner $\sim$3$\times$0.7 degrees of Galaxy in K-band, covering two frequency ranges: 21.20-23.20 GHz and 23.60-25.60 GHz. These frequencies were observed simultaneously using the Compact Array Broadband Backend (CABB), which provides two IF bands, each of 2\,GHz bandwidth. In the CFB 64M-32k mode that was adopted for these observation, each 2\,GHz band was divided into 32 $\times$ 64\,MHz ``continuum'' channels with 16 ``zoombands'' of 2048 $\times$ 32\,kHz channels selected in each IF band for spectral line observations \citep[See][for additional details]{Krieger17}. With this setup, all 6 of the \am\, transitions could be observed simultaneously. The resulting survey data have a $\sim$30'' angular resolution which corresponds to a $\sim$1 pc spatial resolution. 

\subsection{Single-Dish Combination with HOPS and Mopra data}
As the SWAG data do not recover flux on the most extended scales, we combine single dish data to account for the missing short spacings in the interferometer data. We introduce two data sets for the single dish option for this combination. The first data sets consists of the (1,1), (2,2), (3,3), and (6,6) transitions from HOPS (H$_2$O southern Galactic Plane Survey) \citep{HOPS1, Prucell2012,HOPS3} and was taken with the 22-m Mopra telescope.  The HOPS region covered longitudes of 290$\degree$ $<$ $l$ $<$ 360$\degree$ and 0$\degree$ $<$ $l$ $<$ 30$\degree$ and Galactic latitudes of -0.5$\degree$ $<$ $b$ $<$ +0.5$\degree$. Details of the data reduction can be found in \cite{HOPS1}, \cite{HOPS2}, and \cite{HOPS3}.

The second data set (referred to as simply ``Mopra'' for the remainder of the paper) was also taken with the 22-m Mopra telescope and consists of the all transitions from (1,1)-(6,6) transitions. The Mopra Spectrometer (MOPS) digital filterband backend was used for the observations. It comprises an 8.3 GHz total bandwidth, split into four overlapping intermediate frequencies (IFs), each with a width of 2.2\,GHz. It is then possible to simultaneously observe up to 16 spectral windows throughout the full 8.3 GHz bandwidth. Each spectral window consists of 4096 channels, which is equivalent to a velocity resolution of 0.52\,\kms\, at 19.5\,GHz, or 0.37\,\kms\, at 27.5\,GHz, respectively.

The SWAG data were reduced and calibrated \citep{Krieger17} using the MIRIAD package, and we combine the SWAG data with single dish data (HOPS or Mopra) also using MIRIAD. We start by creating a dirty map from the SWAG UV data points and regrid the HOPS/Mopra data sets to the dirty SWAG map. 

Before continuing, we scale the Mopra data to equate the two single dish data sets. Using the (1,1), (2,2), (3,3), and (6,6) transitions we plot the HOPS versus the Mopra data and fit a line to the data. The reciprocal of the slope provides a scaling factor. We average the slopes of all four transitions and find a scaling factor of $\sim$2.238. We apply this scaling factor to scale the Mopra data before the data combination. We can then generate a single dish beam with calculated beam sizes and deconvolve with \textit{mosmem} (30 iterations) to construct a mask that prevent treating regions without significant emission. After restoring the image we create a mask with a 5 sigma cut and deconvolve again using \textit{mosmem}, this time with 100 iterations. We restore the image again to get a combined image. Finally, we merge (using \textit{immerge}) the combined image with the HOPS/Mopra data sets to produce a final cube. 

Because Bania's Clump 2 is outside the SWAG field of view, we do not have the option of using SWAG data. However, we use HOPS data of (1,1), (2,2), (3,3), and (6,6) transitions of \am\, with a resolution of 2$'$ to derive both a warm and a hot temperature component for these sources as well.

\section{Ammonia Spectra and Spectral Fitting}
\label{spectra}

\subsection{Spectra}
\subsubsection{GBT Spectra}
We detect the $(J,K)$ = (8,8) to (11,11) transitions of \am\, toward 16 new positions, shown in Figure \ref{finder} and listed in Table \ref{table:sources}. Four sources consist of only (8,8)-(11,11). For three sources it is due to non-detection of the states higher than (J,K)=(11,11). These three sources all reside in Bania's clump. The fourth source with only (8,8)-(11,11) lines (M1.32-0.13) did not have observations taken due to observation timing constraints. For many of these sources we detect higher transitions up to (14,14). We detect four sources in transitions up to (12,12), seven sources in transitions up to (13,13), and one source in transitions up to the (14,14) line. We also detect (and fit) multiple velocity components in six sources. The spectra are displayed in Figure \ref{fig:gbt_spectra}.

\subsubsection{SWAG Spectra} 
For the 13 sources for which we have SWAG data, we detect all of the $(J,K)$ = (1,1)-(6,6) transitions toward 11 sources. We also detect the (1,1), (2,2), (3,3), (4,4) and (6,6) transitions toward source M1.01+0.02(a) and (1,1)-(5,5) toward source M0.11-0.08. Five sources have multiple velocity components and one source displays resolved hyperfine lines (M0.48-0.01) in the (1,1)-(4,4) transitions. The four sources with two velocity components detected in the GBT data coincide with four of the five sources with two velocity components found in the SWAG data. The spectra for these data can be found in Figure \ref{fig:swag_spectra}.

\subsubsection{HOPS Spectra}
From the HOPS data set we have measurements of the (1,1)-(3,3) and (6,6) transitions for the three sources in Bania's Clump 2. For M3.14+0.41 we find two velocity components which are resolved for (1,1)-(3,3), and for M3.35+0.43 we do not detect the (6,6) line. The HOPS spectra for these sources can be found in Figure \ref{fig:swag_spectra}.

\subsection{Spectra Fitting}
\subsubsection{GBT Spectral Fitting}

\begin{figure*}[h]
\centering
\includegraphics[scale=0.40]{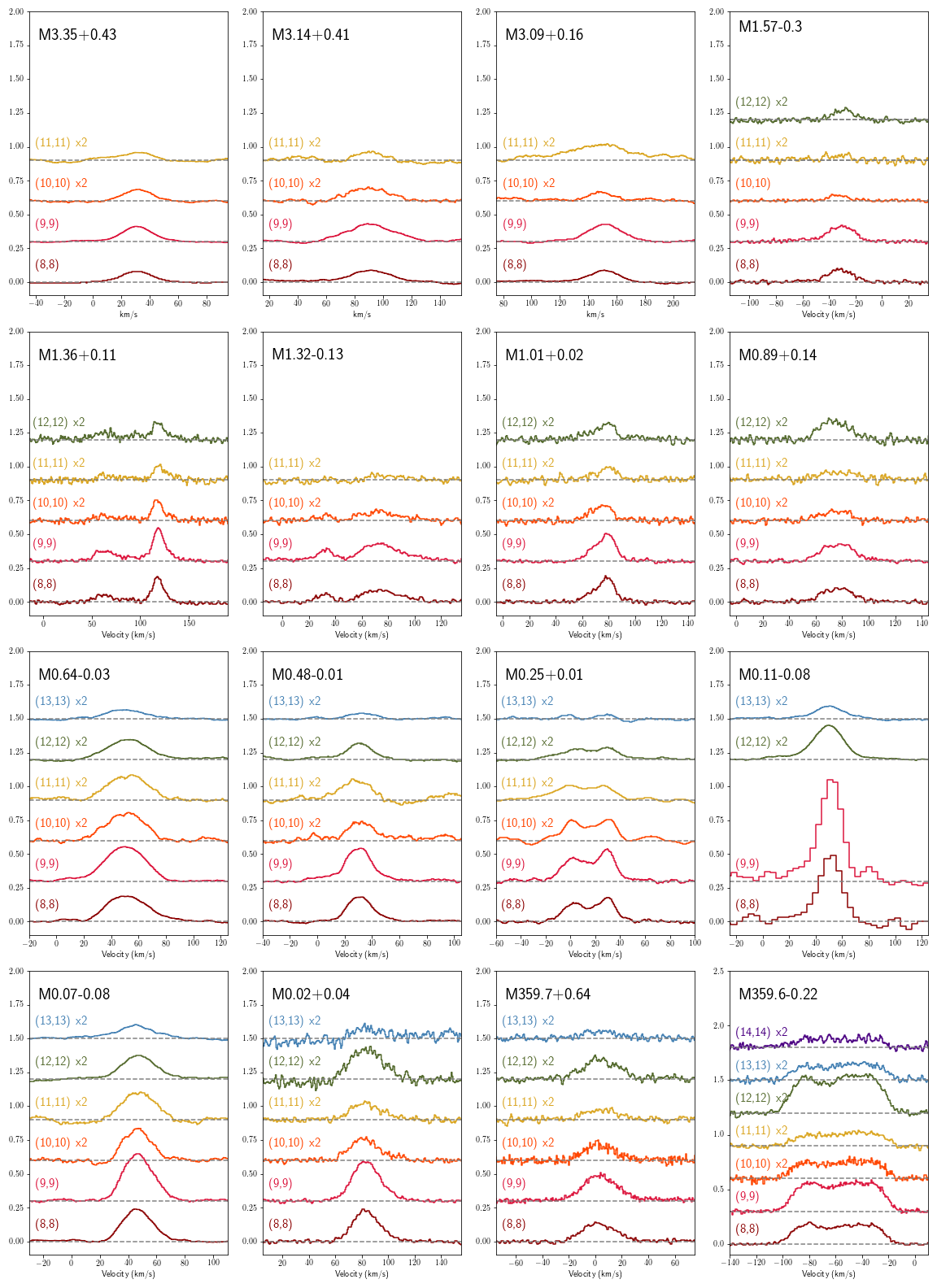}
\caption{Spectra of \am\, for the (8,8)-(14,14) lines in all 16 sources. Lines above and including the (10,10) are multiplied by two to better display characteristics of the lines. These are indicated to the left of each transition line. Lines with emission less than 3$\sigma$ will be used as upper limits.}
\label{fig:gbt_spectra}
\end{figure*}

\begin{figure*}[h]
\centering
\includegraphics[scale=0.40]{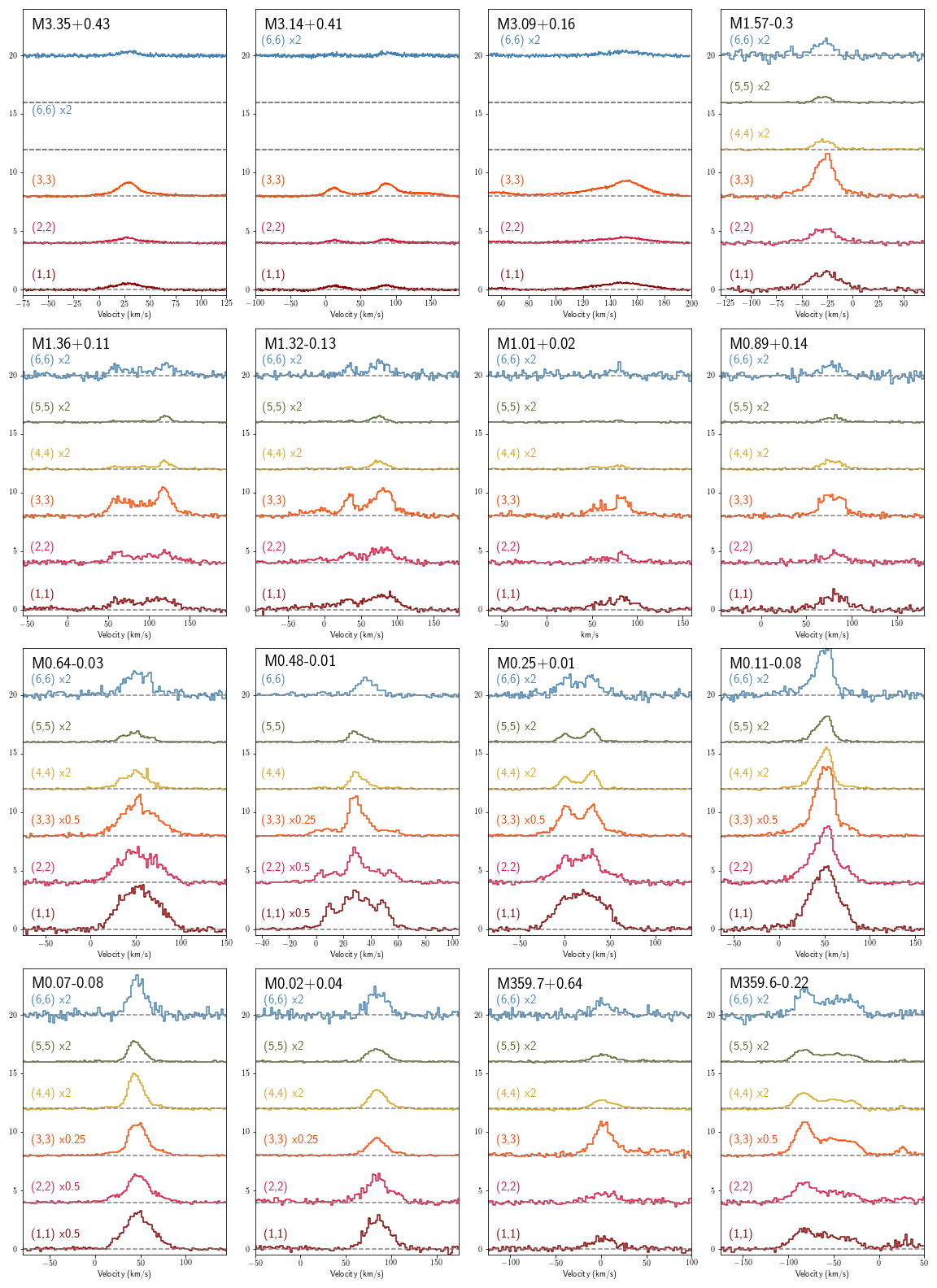}
\caption{Spectra of \am\, for the (1,1)-(6,6) lines in 16 sources. For most sources, lines above and including the (4,4) are multiplied by two to better display characteristics of the lines. Additionally, some sources have been multiplied by 0.5 or 0.25 in order to show full emission without overlapping emission.}
\label{fig:swag_spectra}
\end{figure*}

In fitting the spectra, we use \textit{Pyspeckit} \citep{Ginsburg22-pyspeckit} to obtain parameters including peak intensity, central velocity, and velocity dispersion as well as errors for each parameter. These numbers, and the derived column densities of each transition, can be found in Table \ref{tab:gbt_line_fits} for the 2014 and 2020 GBT data.

Multiple sources exhibit more complex spectra with two velocity components. We fit these spectra with two independent Gaussian components and separately report the line parameters for each component. Sources with multiple velocity components are noted in Table \ref{table:sources}.

For one source (M0.11-0.08) we use a combination of the (8,8) and (9,9) transitions previously observed by \cite{Mills13a} and (12,12) and (13,13) transitions obtained with the 2014 data set. The (8,8) and (9,9) spectra were not observed with VEGAS, the spectrometer used for 2014 observations. This leads to considerably coarser spectral resolution for the (8,8) and (9,9) transitions, which can be seen in the spectra. The GBT spectra for all sources are shown in Figure \ref{fig:gbt_spectra}.

\subsubsection{SWAG \& HOPS Spectral Fitting}
For the SWAG and HOPS data, spectra are extracted from the data cubes in apertures of 26$''$ size for the SWAG data and 2$'$ for the HOPS data. We use the same techniques for spectral fitting for the SWAG and HOPS data as we do for the GBT data. We use \textit{pyspeckit} \citep{Ginsburg22-pyspeckit} to fit the Gaussians by supplying initial guesses for each parameter which guides the fitting. For the one source in which hyperfine structure can be resolved, M0.48-0.01, we use an ammonia fitting template in \textit{pyspeckit} to measure the line opacity from the relative strength of the main line and hyperfine satellite components. This fitting returns an opacity-corrected peak intensity as well as the linewidth and central velocity. For the remaining sources, the hyperfine structure is poorly resolved in the J$\leq$3 lines, thus we fit the data with a Gaussian profile, though this does not not allow us to fit for the opacity and will potentially overestimate the velocity dispersion. For J$>$3, the hyperfine structure is no longer prominent and it is sufficient to fit the lines with a Gaussian profile. The SWAG spectra are shown in Figure \ref{fig:swag_spectra}.

\section{Temperatures}
\label{temperatures}

\subsection{Column Densities}
\label{column_densities}
To derive rotational temperatures from \am\, we first determine the level population for each observed transition. For lines observed in emission and assuming the emission to be optically thin, the column density can be calculated as: 

\begin{equation}
    N(J,K) = \frac{1.55 \times 10^{14} cm^{-2}}{\nu} \frac{J(J+1)}{K^{2}} \int T_{mb}\, dv
    \label{col_density_eqn}
\end{equation}

\noindent as in \cite{Mauers03}. Here the transition frequency, $\nu$, is in units of GHz, and $\int T_{mb} d\nu$ is the integrated intensity of the line, in units of K \kms. Calculated column densities for the GBT data are recorded in Table \ref{tab:gbt_line_fits} along with additional measured line parameters from the Gaussian fits.

For the SWAG data, we assume optically thin emission due to the lack of resolving the hyperfine lines for all sources except for M0.48. Given that opacity typically decreases for higher states, its effect on column density calculations is less for higher J transitions \citep{Krieger17}. Calculated column densities for the SWAG data are recorded in Table \ref{tab:gbt_line_fits} as well as the measured line parameters from the Gaussian fit. 

In M0.48, we fit hyperfine line profiles for the J$\leq$3 lines. Using the measured optical depth, we derive ratios of the ammonia column density N(J,K) and excitation temperatures T$_{ex}$ for the two states of a given inversion doublet using the equation: 

\begin{equation}
    \frac{N(J,K)}{T_{ex}} = 1.61 \times 10^{14} \times \frac{J(J+1)}{K^{2}\nu} \times \tau \times \Delta v_{1/2}
\end{equation}

\noindent from \citep{Huttem95}. Here, $\nu$ is in units of GHz, $v_{1/2}$ is the FWHM line width in \kms, and $\tau$ is the optical depth obtained from profile fitting with \textit{pyspeckit}. Calculated column densities for M0.48 data are displayed in Table \ref{tab:swag_M0.48} with additional line measurement parameters from the fits. \\

\subsection{Rotational Temperatures}
\label{rot_temps}

\begin{figure*}[th!]
\begin{centering}
\includegraphics[width=\textwidth]{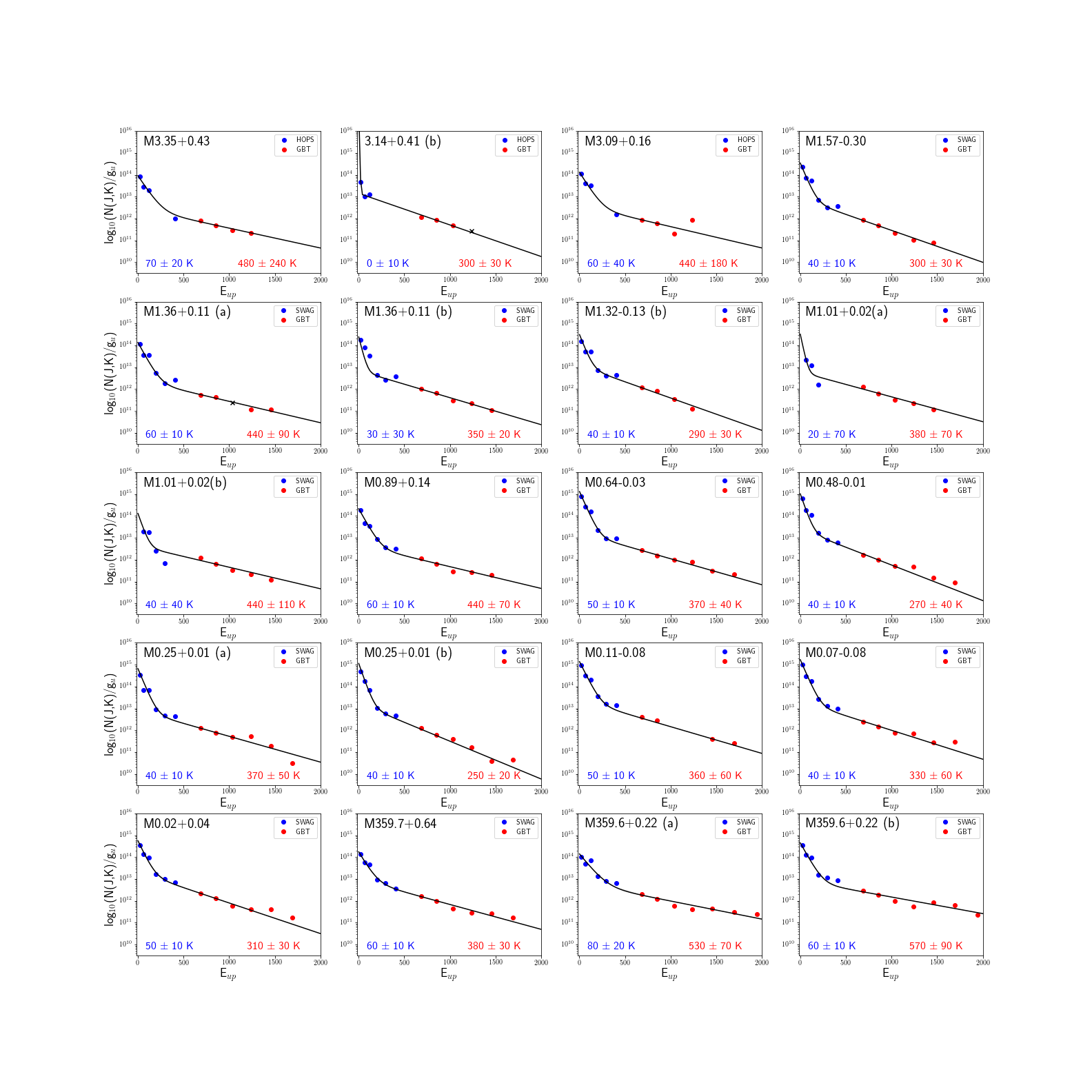}

\caption{Boltzmann diagrams for each velocity component. For sources with two velocity components we denote the separate components as (a) and (b) with (a) being the lower velocity component and (b) being the higher velocity component of the line. From these we derive rotational \am\, temperatures for each velocity component and use these derived temperatures for future calculations. Note that the error bars are smaller than the markers.}

\label{boltz}
\end{centering}
\end{figure*}

At low relative kinetic temperatures \am\, rotational temperatures correspond well to the true gas kinetic temperature. However, for higher gas temperatures, rotational temperatures often considerably underestimate the kinetic temperature \citep{Walmsley83}. The amount by which the kinetic temperature is underestimated is dependent on the upper level energy of the highest transition being used to measure the gas. When the upper level energy is equal to or less than the kinetic temperature, the difference between the rotational and kinetic temperature is decreased. Rotational temperatures measured from higher \am\, transitions will better trace the hot gas components, although they may still underestimate their true kinetic temperature. For temperatures $<200$K, the \am\, thermometer is well-calibrated to account for the difference between the rotational and kinetic temperature, and kinetic temperatures can be derived from the observed rotational temperatures \citep{Walmsley83,Maret09}. However, for higher temperatures, the the rotational temperature is best interpreted as a lower limit on the true kinetic gas temperature. 

To determine the rotational temperature we invoke Boltzmann statistics. For each transition, we consider a contribution to its column density from both a warm and a hot temperature component. While in general column densities in transitions observed with the GBT (J$\geq$8) are dominated by the hot component and transitions observed with SWAG and HOPS (J$\leq$6) are dominated by the warm component, this division does not always apply. We express the total column density from the hot component, N$^{H}$, by equation \ref{N_hot} and the total column density from the cool component, N$^{C}$, by equation \ref{N_cool}

\begin{equation}
    N^{H} = g_{u} N^H_0 \cdot e^{(\frac{-E_{u}}{T_{rot}^{H}})}
    \label{N_hot}
\end{equation}

\begin{equation}
    N^{C} = g_{u} N^C_0 \cdot e^{(\frac{-E_{u}}{T_{rot}^{C}})}
    \label{N_cool}
\end{equation}

\noindent where g$_{(J,K)}$  is the degeneracy factor, E is the energy of  the upper level of a symmetric top, E$_{u}$ = BJ(J+1) + (C-B)K$^2$, T$^H_{rot}$ hot rotational temperature, N$^H$ is the expected column density for the ground state (J,K)=0 for the hot component, and N$^C$ is the expected column density for the ground state (J,K)=0 for the cool component. 

To find the overall total column density for both the hot and warm components we sum the two.

\begin{equation}
    N_{tot} = \sum_{i=1}^{6} g_{u,i} N^H_0 \cdot e^{(\frac{-E_{u,i}}{T_{rot}^{H}})} +   
    \sum_{i=1}^{6} g_{u,i} N^C_0 \cdot e^{(\frac{-E_{u,i}}{T_{rot}^{C}})} 
\end{equation}

From this equation we have four unknowns to fit, T$_{rot}^{C}$, T$_{rot}^{H}$, N$^C$, and N$^H$, for each source. These were jointly fit to all observed transitions.

In total we observe 6 sources with two velocity components. Two velocity components are treated independently resulting in two fits, one for each velocity component, and thus two temperatures for the source. For the GBT data, we observe four sources (M1.36+0.11, M1.32-0.13, M0.25+0.01, and M359.6-0.22) with two velocity components, which also present two velocity components in the SWAG data. In addition, have two sources (M3.14+0.41 and M1.01+0.02) that only exhibit two velocity components in the SWAG data. 

For M1.32 we did not use the first (lower) velocity component due to the (10,10) and (11,11) only being upper limits. For the second component with a higher velocity, we measure (8,8)-(11,11) providing a better fit to the data. M3.14 exhibited two velocity components in the HOPS transitions in three out of four lines, which did not provide a sufficient fit. We used the second velocity component to fit for a hot component. For this source, we do not report a cool component temperature. Additionally, for M1.01+0.02(a) we only fit the lower four lines of the SWAG data due to a non-detection of the lower velocity component in the (5,5) and (6,6). 

For the observed positions we find hot rotational temperatures of $\sim$250-570\,K. The hottest source, M359.6+0.22, displays two velocity components resulting in two of the hottest temperatures found at 530 $\pm$ 70\,K and 570 $\pm$ 90\,K. Four additional sources express temperatures greater than 400\,K (3.35+0.43, 3.09+0.16, M1.36+0.11(a), M1.01+0.02(b), and M0.89+0.14). These sources as spread throughout the Galactic center and two reside in Bania's clump 2 located at a projected galactocentric radius of $\sim$450\,pc. For the cool component we find rotational temperatures of 20-80\,K. The source with the warmest `cool' component, M1.36+0.11(a), is found to have a temperature of 100 $\pm$ 20\,K. This source also has one of the hottest temperatures when investigating the hot component.

\subsection{The Gas Fractions}
To calculate the hot (T\,$>$\,250\,K) and cool (T\,$\lesssim$\,100\,K) gas fractions for each source we need to determine the total column density of cool and hot gas. We do this by finding column density for each (J,K) in the following equation: 

\begin{equation}
    N = g_u N_0 \cdot e^{-E_u/T_{rot}}
    \label{Ng}
\end{equation}

\noindent where E$_u$ is the upper-level energy in K for a given (J,K) state, g$_u$ is the total degeneracy of the observed state, T$_{rot}$ is the rotational temperature measured from the fits of the Boltzmann diagrams for a given component, and I is is the expected column density for the ground state (J,K)=0 for the given component, 

Using our fit parameters, T$_{rot}^{C}$, T$_{rot}^{H}$, N$^C$, and N$^H$, we can calculate N$^C$(J,K) and N$^H$(J,K) for all metastable \am\, levels from $J=0$ to $J=14$. We sum over all the N$_C$ values to give a total cool column density (N$^C_{tot}$). Similarly, we sum over all the N$_H$ values to provide a total hot column density (N$^H_{tot}$). The sum of these two values will result in total column density for all the gas. From the total column density we calculate the hot gas fraction (N$^H_{tot}$/N$_{tot}$) and the cool gas fraction (N$^C_{tot}$/N$_{tot}$).

The hot gas fraction ranges from 9-27\% resulting in a cool gas fraction range of 71-92\%. These fractions can be found in Table \ref{temps_fractions} as well as the ratio of cool to hot gas which ranges from 1.1 to 10.8.

\begin{figure}
    \centering
    \includegraphics[width=0.5\textwidth]{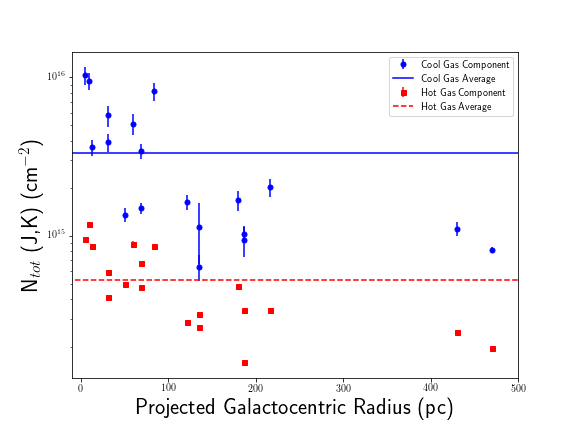}
    \caption{The total gas fraction for the hot gas and warm gas where the hot gas is represented by the blue squares and hot gas is represented by the red circles. The gas fraction for each component is averaged for all sources and indicated on the plot by a solid blue line for the cool component and a dotted red line for the hot component.}
    \label{fig:radius_col_density}
\end{figure}

In Figure \ref{fig:radius_col_density}, we show the total gas fraction for the hot and cool gas 
as a function of projected galactocentric radius for each component to show the overall increase in column density of the cool component to hot component. It is observed that the cool gas fraction is significantly larger than the hot gas fraction, and a strong relationship between galactocentric radius and total hot or total cool column density is not exhibited. We calculate an R$^2$ value of 0.43 for the hot gas component and an R$^2$ of 0.31 for the cool component. However, while we do not find a strong correlation, we do notice both components appear to decrease at the same rate, implying some sort consistency between the hot and cool components. 


\begin{table*}[ht!]
\centering

\caption{Source Temperatures \& Gas Fraction}
\begin{tabular}{lcccccccc}
\hline \hline
{\bf Source} & {\bf T$_{hot}$} & {\bf T$_{cool}$ } & {\bf T$_{H_2CO}$$^{(\mathrm a)}$} & {\bf N$_{hot}$ } & {\bf N$_{cool}$} & {\bf N$_{hot}$/N$_{total}$} & {\bf N$_{warm}$/N$_{total}$} & {\bf N$_{cool}$/N$_{hot}$} \\
{} & {(K)} & {(K)} & (K) & {(10$^{14}$ cm$^{-2}$)} & {(10$^{14}$ cm$^{-2}$)} & {\%} & {\%} &  \\
\hline

{M3.35+0.43}     & {480 $\pm$ 240} & {70 $\pm$ 20} & {-----}  &  {1.9 $\pm$ 0.05} & {8.2 $\pm$ 0.4}  & {19 $\pm$ 1} & {81 $\pm$ 5} & {4.2}\\
{M3.14+0.41(b) } & {300 $\pm$ 30}  &   {0 $\pm$ 0} & {-----}  &  {5.9 $\pm$ 0.04} & {-----}  & {-----} & {-----} & {-----}\\
{M3.09+0.16}     & {440 $\pm$ 180} & {60 $\pm$ 40} & {-----}  & {2.5 $\pm$ 0.06}  & {11.1 $\pm$ 1.2}  & {18 $\pm$ 1.6} & {82 $\pm$ 11} & {4.5}\\
{M1.57-0.30}     & {300 $\pm$ 30}  & {40 $\pm$ 10} & {-----}  &  {3.4 $\pm$ 0.1}   & {20.3 $\pm$ 2.7}  & {15 $\pm$ 1.7} & {86 $\pm$ 15} & {5.9} \\
{M1.36+0.11(a)}  & {440 $\pm$ 90} & {60 $\pm$ 10} & {-----} & {1.6 $\pm$ 0.04}  & {10.2 $\pm$ 1.1}   & {14 $\pm$ 1.3} & {86 $\pm$ 13} & {6.4} \\
{M1.36+0.11(b)}  & {350 $\pm$ 20}  & {30 $\pm$ 30} & {-----}    & {3.4 $\pm$ 0.1}   & {9.5 $\pm$ 2.1}  & {27 $\pm$ 4.3} & {74 $\pm$ 2.0} & {2.8} \\
{M1.32-0.13(b)}  & {290 $\pm$ 30}  & {40 $\pm$ 10} & {100}    & {4.8 $\pm$ 0.2}    & {16.9 $\pm$ 2.4}  & {22 $\pm$ 2.6} & {78 $\pm$ 14} & {3.5} \\
{M1.01+0.02(a)}  & {380 $\pm$ 70}  & {20 $\pm$ 70} & {110}    & {3.2 $\pm$ 0.09}  & {11.4 $\pm$ 4.8}   & {22 $\pm$ 7.3} & {78 $\pm$ 42} & {3.6}\\
{M1.01+0.02(b)}  &  {440 $\pm$ 110} & {40 $\pm$ 40} & {110}    & {2.7 $\pm$ 0.06}   & {6.4 $\pm$ 1.2}  & {30 $\pm$ 4.0} & {71 $\pm$ 16} & {1.1} \\
{M0.89+0.14}     & {440 $\pm$ 70}  & {60 $\pm$ 10} & {-----}  & {2.9 $\pm$ 0.06}    & {16.3 $\pm$ 1.8}  & {15 $\pm$ 1.4} & {85 $\pm$ 12} & {5.7} \\
{M0.64-0.03}     & {370 $\pm$ 40}  & {50 $\pm$ 10} & {110}    & {8.6 $\pm$ 0.2}   & {81.6 $\pm$ 10}  & {9.6 $\pm$ 1.1} & {90 $\pm$ 16} & {9.5}\\
{M0.48-0.01}     & {270 $\pm$ 40}  & {40 $\pm$ 10} & {70}     & {9.0 $\pm$ 0.3}   & {51.1 $\pm$ 0.8}  & {15 $\pm$ 2.0} & {85 $\pm$ 17} & {5.7}\\
{M0.25+0.01(a)}  & {370 $\pm$ 50}  & {40 $\pm$ 10} & {100}    & {4.1 $\pm$ 0.1}    & {39.0 $\pm$ 5.2}  & {9.5 $\pm$ 1.2} & {91 $\pm$ 16} & {9.5}\\
{M0.25+0.01(b)}  & {250 $\pm$ 20}  & {40 $\pm$ 10} & {100}    & {5.9 $\pm$ 0.2}  & {57.7 $\pm$ 8.7}   & {9.3 $\pm$ 1.3} & {91 $\pm$ 19} & {9.8}\\ 
{M0.11-0.08}     & {360 $\pm$ 60}  & {50 $\pm$ 10} & {120}    & {11.9 $\pm$ 0.3}  & {95.0 $\pm$ 11.1}  & {11 $\pm$ 1.2} & {89 $\pm$ 14} & {8.0}\\
{M0.07-0.08}     & {330 $\pm$ 60}  & {40 $\pm$ 10} & {100}    & {9.6 $\pm$ 0.3}   & {104 $\pm$ 13.4}  & {8.5 $\pm$ 1.0} & {92 $\pm$ 16} & {10.8}\\
{M0.02+0.04}     & {310 $\pm$ 30}  & {50 $\pm$ 10} & {110}    & {8.6 $\pm$ 0.3}  &  {36.2 $\pm$ 4.4} & {19 $\pm$ 2.0} & {81 $\pm$ 13} & {4.2} \\
{M359.7+0.64}    & {380 $\pm$ 30}  & {60 $\pm$ 10} & {90}     & {5.0 $\pm$ 0.1}    & {13.6 $\pm$ 1.4}   & {27 $\pm$ 2.2} & {73 $\pm$ 9}& {2.7} \\
{M359.6+0.22(a)} & {530 $\pm$ 70}  & {80 $\pm$ 20} & {140}    & {4.7 $\pm$ 0.01}    & {15.0 $\pm$ 1.2}   & {24 $\pm$ 1.6} & {76 $\pm$ 8} & {3.2} \\
{M359.6+0.22(b)} & {570 $\pm$ 90} & {60 $\pm$ 10} & {140}     & {6.7 $\pm$ 0.1}    & {34.5 $\pm$ 3.7}  & {16 $\pm$ 1.5} & {84 $\pm$ 12} & {5.1} \\

\hline
\end{tabular}

\label{temps_fractions}  
$^{\mathrm a}$ From \cite{Ginsburg16}
\end{table*}

\section{Discussion} 
\label{discussion}

\subsection{Investigating Temperature Correlations}
We detect emission from highly-excited ($J>8$) metastable transitions of \am\, toward 16 new positions in the inner projected galactocentric radius of 500 parsecs of the Milky Way. Rotational temperature fits to these transitions find $T_{rot}$=$\sim$ 250-570\,K. This more substantial sample allows us to investigate potential correlations between the rotational temperature and other measured gas properties, which may help to understand the origin of these highly-excited \am\, lines and the overall heating mechanisms at work for molecular gas in the CMZ. 

\subsubsection{Spatial Distribution of the Highly Excited \texorpdfstring{\am}{Ammonia}}
\begin{figure}[th!]
    \centering
    \includegraphics[width=0.5\textwidth]{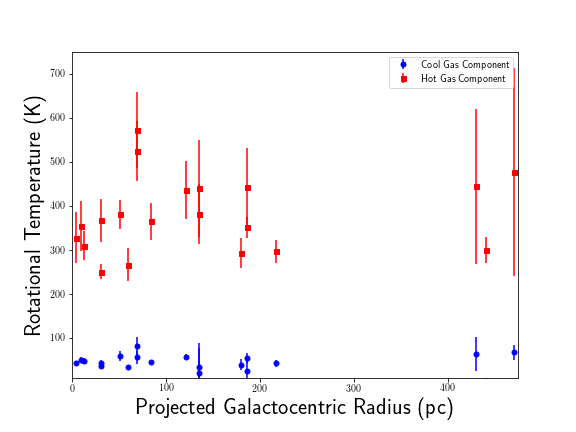}
    \caption{Rotational temperatures plotted versus projected radius (in pc) for each velocity component.}
    \label{fig:radius_temp}
\end{figure}

We compute the minimum possible projected Galactocentric distance for each of the positions for which we observe \am\, spectra, and these distances are reported in Table \ref{table:sources}. Figure \ref{fig:radius_temp} shows the rotational temperatures versus the projection radius (in pc) for both GBT data and SWAG data. It is clear that a correlation between projected radius and rotational temperature is absent.

Our hottest sources, M359.6, produces temperatures of 530$\pm$70\,K and 570$\pm$90\,K (due to two velocity components). M359.6 sits at a galactocentric radius of $\sim$70\,pc, almost 10 times the projected distance of the closest two sources observed, M0.11 and M0.07. M0.11 and M0.07 reside at a projected galactocentric radii of $\sim$10\,pc and $\sim$5\,pc, respectively, with hot component temperatures of 360$\pm$60\,K and 330$\pm$60\,K, respectively. \cite{Mills13a} found temperatures ranging from 350-450K using NH$_3$ transitions (10,10), (11,11), (12,12), (13,13), and (15,15) in 3 molecular clouds within 40\,pc of the central super massive black hole, which is consistent with our findings. 

We report detections of highly excited \am\, up to (J,K)=(11,11) in the Bania's Clump 2, a non-star-forming region on the outskirts of the CMZ. The three sources in Bania's clump are located at a projected galactocentric radius of $\sim$450\,pc, much further than other molecular clouds observed in this study. We derive temperatures for the three Bania's clump sources of $\sim$370-470\,K with all sources having hot or hotter gas than Sgr B2 (370$\pm$40\,K). Sgr B2 is actively forming massive stars and is the densest and most massive giant cloud in the CMZ \citep{Mauers86}. However, we find comparable temperatures in the Bania's Clump 2, which is not nearly so extreme. From the absence of a correlation we infer that the elevated temperatures we find in the CMZ are not galactocentric radius dependent.

\subsubsection{Comparison with H$_2$CO from \cite{Ginsburg16}}
\begin{figure}
    \centering
    \includegraphics[width=0.5\textwidth]{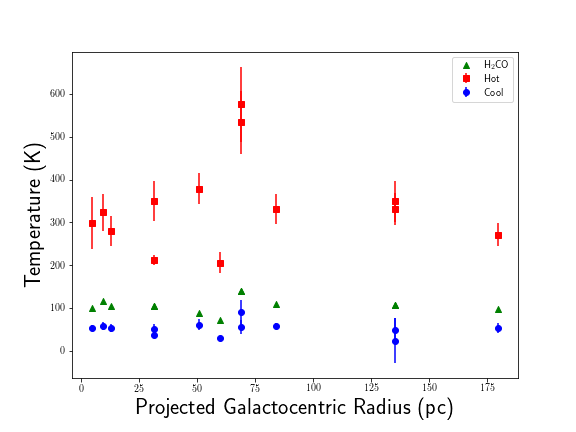}
    \caption{The rotational temperature versus the projected radius displaying temperatures derived from H$_2$CO in the green diamonds (from \cite{Ginsburg16}), the hot component in the red squares, and the cool component in the blue dots. This plot only includes the radius for positions we have overlap and does not cover the full range of galactocentric radius for the entire 16 samples.}
    \label{fig:radius_h2co_temps}
\end{figure}

\cite{Ginsburg16} maps H$_2$CO in the inner 300 pc of the CMZ using the J=3-2 para-formaldehyde transitions with 30'' resolution. They use line ratios to determine the gas temperature of gas with density on the order of 10$^4$-10$^5$ \cm\, and find an effective temperature range of $\sim$40-150\,K with the higher measured temperatures becoming lower limits above $\sim$150\,K due to the energies of the specific transitions used. In Figure \ref{fig:radius_h2co_temps} we see the H$_2$CO derived temperatures consistently lie above the cool temperature component but larger than the `hot' component. Only one of these sources comes close to that 150\,K upper limit, M356.6 with 141.1\,K, but still falls below it. We postulate from this that since the upper limit found for H$_2$CO temperatures ($\sim$150\,K) lie between the hot and cool components, the H$_2$CO must partially be coming from the hot component. These limits could be corroborated better in two ways, (1) by increasing the number samples with overlap in the H$_2$CO data to get a more complete sample and/or (2) by mapping the same region as the H$_2$CO data with lower transitions of ammonia and comparing the cool temperature component derived.

We also looked for trends in the data corresponding to properties of the region (i.e. higher temperatures, star forming regions, high velocity dispersions, quiescent clouds, etc., see Appendix \ref{notes_on_sources}). We find the highest H$_2$CO temperature corresponds to the region with the hottest hot component, M359.6, which is associated with large velocity dispersion. We see some of the more elevated temperatures are associated with some star forming regions, e.g. M0.64, M0.11, and M0.07, but we also see a higher temperature coming from M0.25, which is not known to be a star forming region. M359.7 is also associated with SNR and WR/OB stars, a known source of energy, however, this regions exhibits one of the lowest temperatures. We do not find any obvious trends from these samples. The value of these temperature for each applicable region can be found in Table \ref{temps_fractions}.

\subsubsection{\texorpdfstring{\am\,}{Ammonia} Linewidths}

Galactic center molecular gas is extremely turbulent with line widths (FWHM) of 15-50 \kms\, in comparison to widths of $\sim$1-10 \kms\, typically found in giant molecular clouds in the Galactic disk (e.g. \cite{Bally87}). Line widths in these regions are an important indicator of a dynamical environment.

\begin{table*}[ht!]
\centering
\caption{Source Line Widths}
\begin{tabular}{lcccc}
\hline \hline
\bf{Source} & \bf{Thermal Line Width} & \bf{Non-Thermal Line Width} & \bf{Observed Line Width} & \bf{FWHM}\\
 & (km/s) & (km/s) & (km/s) & (km/s)\\
 
\hline

M3.35+0.43 & 1.29 $\pm$ 0.91 & 6.10 $\pm$ 0.35 & 6.24 $\pm$ 0.32 & 14.70 $\pm$ 0.76 \\
M3.14+0.41 (b) & 1.02 $\pm$ 0.32 & 7.55 $\pm$ 1.06 & 7.63 $\pm$ 1.09 & 17.97 $\pm$ 2.57 \\
M3.09+0.16 & 1.25 $\pm$ 0.78 & 8.23 $\pm$ 0.44 & 8.34 $\pm$ 0.44 & 19.63 $\pm$ 1.04 \\
M1.57-0.30 & 1.02 $\pm$ 0.30 & 6.24 $\pm$ 0.44 & 6.33 $\pm$ 0.46 & 14.91 $\pm$ 1.08 \\
M1.36+0.11 (a) & 1.24 $\pm$ 0.55 & 10.99 $\pm$ 1.40 & 11.07 $\pm$ 1.43 & 26.07 $\pm$ 3.36 \\
M1.36+0.11 (b) & 1.11 $\pm$ 0.28 & 4.60 $\pm$ 0.23 & 4.75 $\pm$ 0.24 & 11.18 $\pm$ 0.57 \\
M1.32-0.13 (b) & 1.01 $\pm$ 0.34 & 8.02 $\pm$ 0.48 & 8.09 $\pm$ 0.49 & 19.05 $\pm$ 1.15 \\
M1.01+0.02(a) & 1.15 $\pm$ 0.49 & 5.73 $\pm$ 0.23 & 5.86 $\pm$ 0.22 & 13.80 $\pm$ 0.52 \\
M1.01+0.02(b) & 1.24 $\pm$ 0.62 & 5.72 $\pm$ 0.24 & 5.86 $\pm$ 0.22 & 13.80 $\pm$ 0.52 \\
M0.89+0.14 & 1.23 $\pm$ 0.48 & 8.61 $\pm$ 0.36 & 8.70 $\pm$ 0.37 & 20.49 $\pm$ 0.86 \\
M0.64-0.03 & 1.13 $\pm$ 0.38 & 12.24 $\pm$ 0.35 & 12.30 $\pm$ 0.35 & 28.96 $\pm$ 0.83 \\
M0.48-0.01 & 0.96 $\pm$ 0.36 & 8.52 $\pm$ 0.27 & 8.58 $\pm$ 0.28 & 20.21 $\pm$ 0.65 \\
M0.25+0.01 (a) & 1.13 $\pm$ 0.41 & 10.05 $\pm$ 0.48 & 10.12 $\pm$ 0.48 & 23.84 $\pm$ 1.14 \\
M0.25+0.01 (b) & 0.94 $\pm$ 0.24 & 5.75 $\pm$ 0.34 & 5.83 $\pm$ 0.35 & 13.73 $\pm$ 0.82 \\
M0.11-0.08 & 1.11 $\pm$ 0.44 & 5.82 $\pm$ 0.19 & 5.93 $\pm$ 0.18 & 13.97 $\pm$ 0.42 \\
M0.07-0.08 & 1.07 $\pm$ 0.45 & 9.63 $\pm$ 0.21 & 9.70 $\pm$ 0.21 & 22.84 $\pm$ 0.49 \\
M0.02+0.04 & 1.04 $\pm$ 0.34 & 9.20 $\pm$ 0.49 & 9.27 $\pm$ 0.49 & 21.82 $\pm$ 1.17 \\
M359.7+0.64 & 1.15 $\pm$ 0.34 & 10.95 $\pm$ 0.60 & 11.02 $\pm$ 0.61 & 25.95 $\pm$ 1.44 \\
M359.6+0.22 (a) & 1.35 $\pm$ 0.49 & 12.13 $\pm$ 0.70 & 12.21 $\pm$ 0.71 & 28.75 $\pm$ 1.67 \\
M359.6+0.22 (b) & 1.41 $\pm$ 0.55 & 15.45 $\pm$ 0.76 & 15.52 $\pm$ 0.77 & 36.55 $\pm$ 1.81 \\

\hline
\end{tabular}
\label{linewidths}         
\end{table*}

For this discussion on line widths we focus on the average line widths for the GBT data only. This is due to the unfit hyperfine lines. Our gaussian fits will result in a broader line widths because of the unresolved hyperfines, but these should be negligible for the GBT data. For GBT data we report line widths (defined at the 1-D velocity dispersion, $\sigma$) of $\sim$ 4-15\kms, corresponding to a full-widths half maximum (FWHM) of $\sim$11-37 \kms, for each component regardless of whether they contain one or two velocity components. The FWHM for each sources and transition is provided in Table \ref{tab:gbt_line_fits}. These line widths are consistent with findings of line widths larger than those found in the disk \citep{Bally88, Morris96, Mills13a}. We find that lower (J,K) transitions, (1,1)-(6,6), exhibit slightly larger line widths as compared to the higher (J,K) values. We find line widths of $\sim$ 5-20 \kms, corresponding to a FWHM of $\sim$11-45 \kms, for the lower (J,K) transitions.

We compare three different line widths, the observed line width, the thermal line widths and the non-thermal line width. We obtain the observed line widths from fitting each transition using \textit{pyspeckit} and averaging the transition values for each source. We find observed line widths between 5-15 \kms\, for the GBT data. The values for each of these line widths can be found in Table \ref{linewidths}.

We then further investigate how these observed line widths are distributed into thermal and non-thermal components. For molecular thermal velocity dispersion we estimate by the following:

\begin{equation}
    \sigma_{th} = \sqrt{(k_B T)/(\mu m_H)}
\end{equation}

\noindent where $\mu$ = $m/m_H$ is the molecular weight, $m$ is the molecular mass, $m_H$ is the proton mass, $k_B$ is the Boltzmann constant, and T is the gas temperature derived from the Boltzmann diagrams. Even for the hot GBT temperatures, thermal line widths (0.96-1.48\kms) are negligible compared to the observed line width.   

\begin{figure}
    \centering
    \includegraphics[width=0.5\textwidth]{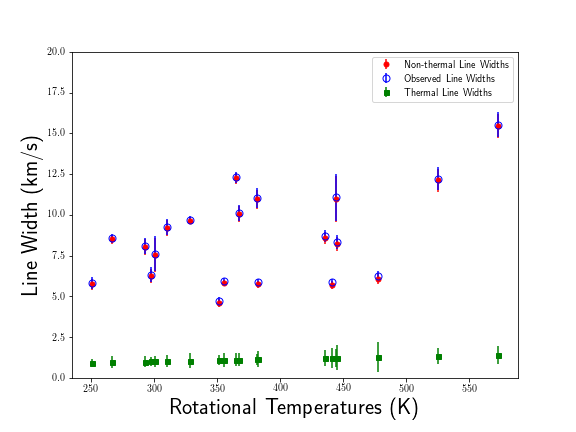}
    \caption{For the GBT data we have plotted the thermal and non-thermal line widths calculated from the rotational temperatures and observed line widths. The green squares indicated the thermal line widths, filled red circles are the non-thermal line widths, and the open blue circles are the observed line widths.}
    \label{fig:gbt_temps_linewidths}
\end{figure}

We derive the non-thermal velocity dispersion using the following equation:

\begin{equation}
    \sigma_{nt} = \sqrt{(\sigma_{obs}^2 - \Delta \nu_{ch}^2/ (2\sqrt{2ln(2)})^2) - \sigma_{th}^2}
\end{equation}

\noindent where $\Delta_{ch}$ and $\sigma_{th}$ are the velocity channel width and thermal velocity dispersion, respectively. If the gas is hotter due to non-thermal effects (i.e shocks, turbulence, cosmic rays, etc.) then we would expect to see broader line widths in regions with stronger shocks, are more turbulent, or are more energetic. Indeed, this is observed in Figure \ref{fig:gbt_temps_linewidths}. With the thermal line widths being so small, we should expect the non-thermal line widths to be approximately equal to the observed line widths. We find non-thermal line widths of $\sim$4-15 \kms\, for the GBT data, approximately the same as the observed line widths. 

\begin{figure}
    \centering
    \includegraphics[width=0.5\textwidth]{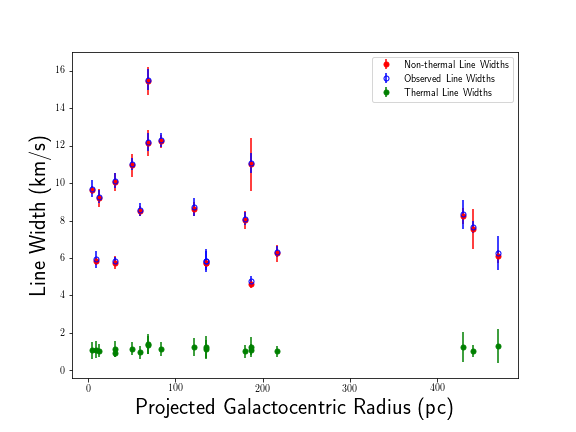}
    \caption{For the GBT data we have plotted the thermal, non-thermal, and observed linewidths versus the projected radius (in pc). The green squares indicated the thermal linewidths, filled red circles are the non-thermal line widths, and the open blue circles are the observed line widths.}
    \label{fig:gbt_linewidth_radius}
\end{figure}

We can conclude four main points from Figure \ref{fig:gbt_temps_linewidths} and Figure \ref{fig:gbt_linewidth_radius}. First, we notice a lack of correlation between the temperature and observed line widths. We calculate an R$^2$ value of 0.383 for the GBT data. This would indicate that temperature and line widths are not well correlated. 

Looking at the hottest sources, M359.6 (530$\pm$70 and 580$\pm$90\,K), M1.36(a) (520$\pm$140), and M3.14(b) (470$\pm$140), we compare their temperatures to the observed line widths. We report line widths of 12.2$\pm$0.707\,\kms\, and 15.5$\pm$7.7\,\kms\, for M3.59.6, 6.3$\pm$0.46\,\kms\, for M3.14, and 8.09$\pm$4.9\,\kms\, for M1.36(a). While these sources are presenting the highest temperatures, they do not present the largest line widths. M359.6 exhibits both the highest temperatures and the largest observed line widths, but this is not consistent for the remainder of sources. We cannot conclude a relationship between the temperatures and linewidths, which implies that temperature is not a main contributor to the broad line widths. 

Second, we find that the observed line widths and the non-thermal linewidths lie almost directly on top of each other in Figure \ref{fig:gbt_temps_linewidths}. We conclude that the line widths are non-thermal suggesting non-thermal motions are dominant over thermal.

Thirdly, we calculate the Mach number for each region using $M$ = $\sqrt{3}\sigma_{nt}/c_s$. The thermal velocity dispersion (sound speed $c_s$) of the particle of mean mass was estimated assuming a mean mass of 2.8 \citep{Kauffmann08}.The derived Mach numbers range from 7-18 for the GBT data, clearly supersonic. We find the largest Mach numbers for the hot component to be from M1.01 with Mach numbers of 19.75$\pm$0.49 and 17.44$\pm$0.40. This source is not co-located with a potential feedback source. M0.02 was the second largest Mach number of 18.89$\pm$0.94s and is co-located with WR/OB stars and a SNR. The third highest Mach number came from M0.11 with a Mach number of 16.51$\pm$0.70. This source is considered to be a young massive star forming site and a candidate for star formation induced by cloud-cloud collision. These three sources experience similar, high Mach numbers despite their different environments. With a larger sample it is possible to better investigate the relationship between various types of environments and Mach numbers. When investigating the temperatures and Mach number, we do not find a correlation between the two parameters.

Lastly, we look into the possible relationship between the line widths and galactocentric radius. In Figure \ref{fig:gbt_linewidth_radius} we show the three different line widths (thermal, non-thermal, and observed) with respect to the galactocentric radius for the GBT transitions and do not see a correlation between the line widths and radius or temperature.

\subsection{Interpreting the Highly-Excited \texorpdfstring{\am\,}{Ammonia} Emission}

The temperature structure of CMZ gas has led to questions about the reliability of \am\, rotational temperatures to trace the gas kinetic temperature in this environment. In general, rotational temperatures measured using \am\, are considered to be reliable estimates of the gas kinetic temperature for low temperatures. The use of \am\, as a kinetic temperature probe (`the ammonia thermometer') has been carefully calibrated and its robustness verified for a number of decades for low temperatures \citep{Danby88,Maret09,Bouhafs17}. Due to the lack of data required to model radiative transfer of the high (J,K) values and determine its affect on the upper transitions, we are unable to deduce accurate T$_{kin}$ values from the very hot gas. Rotational temperatures derived with higher (J,K) transitions should be considered lower limits of T$_{kin}$ for the hot gas.

\subsubsection{Non-thermal Effects}
Formation pumping can occur when molecules are formed via exothermic reactions. Excess energy is distributed between reaction products, and, in principle, is available to excite internal energy states of the molecule. For symmetric top molecules like \am\,, not all of these high-energy states can decay radiatively; de-excitation will feed into highly excited metastable states, for which radiative transitions are forbidden. This can lead to an excess population in these states if collisions are too slow to thermalized them. \cite{Lis12} observes high rotational transitions of H$_3$O$^+$ in Sgr B2(N). One theory they investigate is whether the high rotational temperatures are due to formation pumping in molecular gas irradiated by x-rays emitted by the Galactic center black hole. They postulate \am\, would have to be explained by the same model because \am\, is chemically more stable than H$_3$O$^+$ and would then have more time to relax through collisions and should thus display a lower rotational temperature. When compared with \am, \cite{Lis12} reports H$_3$O$^+$ does not display the higher rotational temperature that we see with \am. Populating the excited levels gives rise to higher rotational temperature. 

As mentioned in section \ref{rot_temps}, rotational temperature underestimate kinetic temperature, therefore, if rotational temperatures are being elevated non-thermally, it is difficult to determine the true kinetic temperatures. We assume \am\, is in thermal equilibrium with \htwo\, gas via collisions given its critical density. If \am\, is being excited to higher rotational states due to additional energy from the reaction, not due to collisions with \htwo\,, temperatures will not be representative of the \htwo\, gas and will overestimate the temperature. Our temperatures are consistent with previous findings such as \cite{Mills13a}, and we can conclude that if formation pumping is occurring, it is consistently throughout the nucleus.

Observations of very highly excited \am\, transitions are not unique to the CMZ. \am\, (6,6)-(14,14) lines, observationally tracing gas with temperatures of 300-450\,K, have been observed around high mass protostars \citep{Goddi2015}. \am\, (9,9) emission, corresponding to E$_u$ nearly 1000\,K, has also been detected in infrared luminous and star forming galaxies on much larger scales ($\sim500$ pc) \citep{Mauers03,Mangum13}. However, the CMZ is unique both in the extremely highly-excited transitions that have been observed \citep[up to $J,K$=18,18 toward the Sgr B2 cloud;][]{Wilson06} and in that the very hot gas traced by these lines is widespread, and not just localized to individual protostars.

\subsection{Distinguishing Between Potential Heating Mechanisms}
The widespread hot temperatures throughout the CMZ favors a global heating mechanism. This heating mechanism must be a consistent source of energy to observe the high rotational transitions. Without a extended and continuous source of energy, molecules will not populate the high metastable states either long enough or at a density great enough for detection over the wide range observed.
 
Many studies agree that gas temperatures reach higher values in the CMZ than seen in the disk of the Galaxy, and that the gas temperatures exceed the dust temperatures. Thus, a global heating mechanism must also be consistent with reduced dust temperatures. Additionally, this global heating mechanism must produce an extended hot component for the denser gas. H$_2$CO traces higher density gas than \am ($>$ 10$^5$ \cm). With the denser gas temperature (derived from H$_2$CO) exceeding the cool component temperatures, we conclude that the extended hot component we observe with ammonia is also true for the denser gas. Therefore, we have two constraints that must be satisfied by a global heating mechanism, (1) reduced dust temperature despite the higher temperatures of the gas and (2) the extended hot component is also true for the denser gas. 

There are variety of possible mechanisms we will explore through the rest of the paper: Photo-dissociation region (PDRs), gravitational heating \citep{GL78}, X-rays \citep{Maloney96}, cosmic rays \citep{GL78,Ao13},  and turbulence \citep{Ao13, Ginsburg16, Goic13,Immer16}.

PDR heating is not believed to be the dominant heating mechanism in the CMZ because while it can explain the discrepancy between the dust and gas temperatures, heating by UV photons in PDRs can only account for 10-30\% of the total \htwo\, column density with a temperature of $\sim$150\,K \citep{RF04}. In addition, \cite{RF01} have measured a warm \am\, abundance in the GC clouds of $\sim$ 10$^{-7}$. This large \am\, abundance is difficult to explain in the context of a PDR since ammonia is easily photo-dissociated by ultraviolet radiation. Thus, other processes should be investigated that are capable of directing heating the gas from the outside \citep{Ao13}. 

While UV radiation only penetrates the outer layers of a region, X-rays are capable of deeply penetrating molecular clouds and heating large quantities of gas. The column density of warm gas in the XDRs can be 10 times higher than that of the PDR, thus XDRs could, in theory, heat large amounts of gas such as those measure in the CMZ \citep{RF04}. However, it has been found that the typical X-ray flux in the CMZ, even allowing for flaring activity of Sgr A*, is insufficient for reproducing the observed high temperatures \citep{RF04, Ao13}. Thus, XDRs do not seem to be a possible heating mechanism in large scales for the bulk of the gas. 

Another possibly source of heating comes from cosmic ray particles being injected into the interstellar medium, presumably by supernova explosions. Star formation activity peaks in the central region of the Galaxy, which increases the amount of energy being injected into this region. Heating by cosmic rays is a pervasive mechanism because cosmic rays can penetrate and travel through a molecular cloud. This suggests that if the cosmic ray ionization rate is high enough, most of the gas in the CMZ should present at the same elevated temperature \citep{RF04}. 

However, \cite{RF01} found that only 30\% of the gas is warm. Thus, \cite{RF04} concluded cosmic rays cannot be heating the bulk of the 150\,K gas. \cite{Ginsburg16}, on the other hand, reported that cosmic rays could still contribute to the heating of the molecular gas, but would not be the dominant heating mechanism and/or the cosmic rays are not uniform across the CMZ. If cosmic rays play an important role in heating the gas, \cite{Ao13} finds that a cosmic-ray ionization rate of at least 1-2 $\times$ 10$^{14}$\,s$^{-1}$ is required to explain the their observed temperatures in the Galactic center. In Sgr B2, however, it is inferred by \cite{YZ07} to have an enhanced flux of cosmic-ray electrons, which is interpreted as the main molecular gas heating source in this region. 

We find cosmic ray ionization rates to be 8.38$\times 10^{-12}$ to 1.02$\times 10^{-11}$\,s$^{-1}$ for the hot temperature component and 3.08$\times 10^{-15}$ to 3.68$\times 10^{-13}$\,s$^{-1}$ for the cool temperature component using the following equation: 

\begin{equation*}
\begin{centering}
    \xi_{CR} = \left[  (12 \frac{dv}{dr} T^{3/2}+4\times10^{-4}n^{1.5})^2 - 16\times10^{-8}n^3 \right] \frac{10^{-17}}{768n^{1/2}\frac{dv}{dr}}
\end{centering}
\end{equation*}

\noindent where T is the derived rotational temperature, n is the density which we assume to be 10$^{4}$ \cm, and $\frac{dv}{dr}$ is the velocity gradient. This equation is a re-arranged version of equation 14 in \cite{Ao13} and used the assumption that the heating is dominated by cosmic rays. We then use our derived rotational temperature to solve for the cosmic ray ionization rate. Because the density is likely variable across the Galactic center, we adopt a density of 10$^3$ and 10$^5$ \cm\, to investigate any dramatic changes. We find no significant difference between the cosmic ray ionization rates when using densities of 10$^3$, 10$^4$, and 10$^5$ \cm. 

\cite{Ao13} and \cite{Ginsburg16} find a cosmic ray ionization rates of at least 1-2$\times$10$^{-14}$ is required in diffuse molecular gas to explain their temperatures. This is also found outside out galaxy in NGC 253 \citep{Harada21}. \cite{Harada21} finds a cosmic ray ionization rate of $>$10$^{-14}$ in order to reproduce observed results. For the cool component, we find consistency with \cite{Ao13} and \cite{Ginsburg16}. However, our required cosmic ray ionization rates found for the hot component are larger by about 1-3 orders of magnitude. Thus far, cosmic rays have not been reported for temperatures exceeding $\sim$200\,K in the Galactic center indicating that our comparison of cosmic ray ionization rates across various temperature components may not be accurate. It is also possible that the high ionization rates are due to the assumption that cosmic rays are the dominant heating source. If this assumption does not hold, our ionization rates could be lower than what has been calculated. We will address this point below. 

We find the cosmic ray heating rate for each source using the following equation: 

\begin{equation}
    \Gamma_{CR} \sim 3.2 \times 10^{-28}\, n\, \left( \frac{\xi_{CR}}{10^{-17 }\,\mathrm{s}^{-1}} \right)\, \mathrm{erg}\, \mathrm{cm}^{-3}\, \mathrm{s}^{-1}
\end{equation}

Our values range from 9.22$\times$10$^{-19}$ to 1.77$\times10^{-17}$ erg cm$^{-3}$ s$^{-1}$ for the hot component and 9.89$\times$ 10$^{-22}$ to 1.18$\times10^{-19}$ erg cm$^{-3}$ s$^{-1}$ for the cool component. We will later compare these to the turbulent heating rates. 

Sources M0.89, M0.02, and M359.7 are associated with known SNR and WR/OB stars. Here we would expect larger cosmic ray ionization, which could increase temperatures in these regions. We find these positions exhibit temperatures of 440$\pm$60\,K, 280$\pm$40\,K, and 380$\pm$40\,K, respectively, for the hot component. We do not find signatures of higher temperature being associated with regions of known SNR or WR/OB stars. This begs the question of whether cosmic rays are contributing significantly to the increased temperature or if another mechanism is dominating.

The broadening of emission lines is due to turbulence, which has been recently suggested to be the most likely candidate for increasing temperatures (\cite{Ao13,Immer16, Ginsburg16, Krieger17}). However, turbulence itself is not an energy source: some other mechanism must then be identified as ultimately driving both the turbulence and the heating in this region. An increase in energy produces more collisions and interactions between molecules, and these interactions contribute to the increase in temperature. Large line width of molecular lines observed toward the CMZ implied the presence of turbulence. We find linewidths (FWHM) ranging from 10-45 \kms, much higher than that observed in the disk, insinuating the presence of turbulence. We also find a slight (positive) correlation between the linewidths and rotational temperatures for the higher (J,K) transitions implying that regions with greater temperature may be more affected by the turbulence. Lastly, we note that the three sources with the greatest linewidths correspond to the sources with two velocity components. Because each velocity component was treated as a source, this is not due to the sum of two components, but rather may be due to the influence of turbulence on the gas in these regions.

We compute average turbulent heating rates from our derived temperature, an estimated \htwo\, density of n=10$^{4}$ cm $^{-3}$, the 1-D velocity dispersion in units of \kms, and the cloud size L in units of pc using the following equation: 

\begin{equation}
    \Gamma_{Turb} = 3.3 \times 10^{-27}\, n\, \sigma^3 L^{-1}\,\, erg\, cm^{-3}\, s^{-1}
\end{equation}

\noindent Here, we have chosen the cloud size to be the size of the beam, which is $\sim$0.033\,pc. 
We derive turbulent heating rates of 1.06$\times$ 10$^{-19}$ to 3.7$\times$ 10$^{-18}$ erg cm$^{-3}$ s$^{-1}$ for the hot component and 1.3$\times$ 10$^{-19}$ to 7.1$\times$ 10$^{-18}$ erg cm$^{-3}$ s$^{-1}$ for the cool. 

We find for the hot component that the heating rates are approximate the same for the cosmic ray heating rate and the turbulent heating rate. Again, because the density of the Galactic center varies, we test various densities (10$^3$, 10$^4$, and 10$^5$ \cm) and find that the turbulent heating rates vary about about an order of magnitude from each other. 

\begin{figure}[h!]
    \centering
    \includegraphics[width=0.5\textwidth]{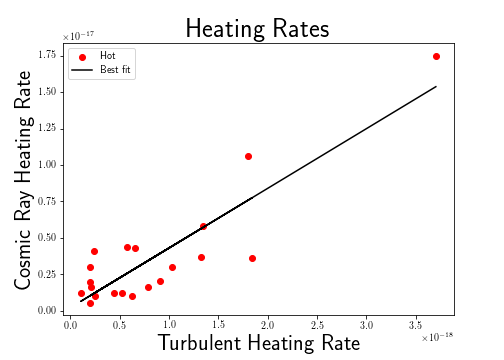}
    \caption{Turbulent heating rates versus cosmic ray heating rates for the GBT data with an R$^2$ value of 0.77.}
    \label{fig:turb_cr_heating_rates}
\end{figure}

For the cool component we find the turbulent heating rates to be 2-3 orders of magnitude larger than the cosmic ray heating rates. \cite{Ao13} and \cite{Ginsburg16} both argue that it is unlikely that that cosmic rays are the dominant heating mechanism or that they are less important in turbulent sub-regions of the CMZ. In Figure \ref{fig:turb_cr_heating_rates} we can see a linear relationship (with an R$^2$ value of 0.77 and a slope of 4.52) between the cosmic ray heating rate and the turbulent heating rate. With this linear relationship we argue that both cosmic rays and turbulence are more than likely required to heat the gas and that the degree to which one dominates may vary depending on the hot temperatures detected. 

\section{Conclusions}
Based on our detection of NH$_3$, we present the following findings
    \begin{enumerate}
        \item We find consistent two temperature fits for the data, the warm temperature component around 60-100\,K and the hot component around 250-570\,K. These temperatures have no correlation with the galactocentric radius regardless of the component. High temperatures detected in Bania's clump 2 are some of the hottest temperature we detect located at a projected radius of $\sim$450\,pc. 
        
        \item We see a lack of correlation between star formation and temperature. Bania's clump 2 is known to not be an active star formation region, but contains some of the hottest temperatures in this sample. On the other hand, regions like Sgr B2, Sgr B1, 20\kms\, cloud and the 50 \kms\, cloud are all well known star forming regions, but do not exhibit temperature hotter than their non-star forming counterparts.
        
        \item The line widths exhibited in these sources are almost entirely non-thermal, suggesting non-thermal motions are responsible for the elevated temperature we observe. From the calculated line widths we find associated Mach number that clearly indicate that the gas is supersonic or highly supersonic.
        
        \item After investigating heating rates of cosmic rays and turbulence we conclude that both turbulence and cosmic rays are needed to heat the gas, but the degree to which one dominates is still unresolved.
    \end{enumerate}
    
\section{Acknowledgements}
We wish to thank the Green Bank Observatory staff and, especially Amber Bonsall and Dave Frayer for their helpful discussion with regards to data calibration. We wish to thank the National Astronomy Consortium (NAC) for their continuous support and the National Radio Astronomy Observatory (NRAO) for funding associated with REUs and the Grote Reber Fellowship.

\appendix

\section{Notes on Sources}
\label{notes_on_sources}

\begin{flushleft}

\textbf{\textit{Bania's Clump 2}}: \textit{M3.35+0.43, M3.14+0.41, M3.09+0.16} \\
\vspace{0.5cm}

Three pointings reside in Bania's clump 2 at a galactocentric radius of $\sim$450\,pc. This region is not known to be a star forming region. The molecular gas is distributed in a complex of 16 individual CS emitting cores, each having densities in excess of 2$\times$ 10$^4$ \cm\, and masses greater than 5$\times$ 10$^5$ M$_\odot$ \cite{Bania86}. These three sources correspond to the location of three CS cores described by \citep{Bania86}. Bania's clump 2 is not known to be a region of active star formation.  \\

\end{flushleft}

\begin{flushleft}
\textbf{\textit{ALMA Sources}}: \textit{M1.57-0.3, M1.32-0.13, 1.36+0.11, M1.01+0.02, M0.89+0.14, M0.02+0.04, M359.7+0.64, 
M359.6-0.22}\\
\vspace{0.5cm}

\noindent The ALMA sources were determined based on two criteria: (1) the temperature in a 2,2/4,4 ratio map of ammonia must be at least 1.5x the surrounding temperature, and (2) We chose regions with energetic feedback sources near or large velocity dispersion from a moment 2 map from the ammonia (3,3) line. The following are the breakdown of the regions that are co-located with potential feedback sources. \\

\end{flushleft}

\begin{enumerate}
    \item M1.57-0.3: Uncertain
    \item M1.32-0.13: Dynamical (large $\Delta$v)
    \item M1.36+0.11: Dynamical (large $\Delta$v)
    \item M1.01+0.02: Uncertain
    \item M0.89+0.14: SNR
    \item M0.02+0.04: SNR and WR/OB stars
    \item M359.7+0.64: WR/OB stars
    \item M359.6-0.22: Dynamical (large $\Delta$v)
\end{enumerate}

\begin{flushleft}
\textbf{\textit{Giant Molecular Clouds}}: \textit{M0.64-0.03, M0.48-0.01, M0.25+0.01, M0.07-0.08, M0.11-0.08}\\
\vspace{0.5cm}

\noindent \textit{M0.64-0.03}: This source is in the Sgr B2 complex, specifically Sgr B2-West. Sgr B2 is one of the most massive star forming region in the galaxy with a mass of 10$^7$\,M$\odot$. Sgr B2 has a higher density ($>$ 10$^5$ \cm\,) and dust temperature ($\sim$50-70\,K) compared to the other star forming region in the Galactic plane (\cite{Schmiedeke16}). \\
\vspace{0.5cm}

\noindent \textit{M0.48-0.01}: The source is known as Sgr B1 which is located southeast of the Sgr B2 complex. This region is a massive star forming region dominate by extended features rather than compact ones. However, it is not absent to compact sources and consist of five compact HII regions. This would suggests that although Sgr B1 may be evolved, the process of star formation is continuing presently (\cite{Mehringer93}). \\ 
\vspace{0.5cm}

\noindent \textit{M0.25+0.01}: The Brick is associated with the Dust Ridge. Despite the substantial reservoir of dense material ($>$10$^5\odot$ of material within a radius of only a few parsec), no evidence of embedded star formation has been observed other than a water maser that coincides with a compact millimeter continuum source (\cite{Lis94,Immer2012,Kauffmann13,Mills2015,Lu19}). The lack of on-going star formation in the Brick coupled with similar evidence in other CMZ clouds, has been argued to favour an environmentally-dependent critical density threshold for star formation (e.g. \cite{Kruijssen14,Walker18,Ginsburg18,Barnes2019}). \\
\vspace{0.5cm}

\noindent \textit{M0.07-0.08}: The 20 km/s cloud is a massive molecular cloud with extremely turbulent regions with large linewidths ($>$10\kms). This cloud presents signatures of star formation and may be triggered by a tidal compression as it approached the pericenter (\cite{Lu15,Lu17}). \\
\vspace{0.5cm}

\noindent \textit{M0.11-0.08}: This cloud, the 50 km/s cloud, is in close proximity to the 20 km/s cloud and is one of the bright molecular clouds in molecular emission lines in the Sgr A region. The cloud is  considered to be a young massive star forming site and a candidate for star formation induced by cloud-cloud collision (\cite{Uehara19}). \\
\vspace{0.5cm}

\end{flushleft}

\clearpage

\centering
\begin{longtable*}{l l c c c c c}
\caption{Measured Line Parameters and Column Densities From GBT}
\\
\hline
Source & Transition & v$_{\mathrm{cen}}$ & v$_{\mathrm{fwhm}}$ & Peak T$_{\mathrm{MB}}$ & $\int T_{\mathrm{MB}} dv$ & N$_u$  \\
        &            & (km s$^{-1}$)      & (km s$^{-1}$)       & (K)     &  (K km s$^{-1}$)    &   ( 10$^{13}$ cm$^{-2}$)  \\
\hline
\hline
\endfirsthead
\caption{Measured Line Parameters and Column Densities From GBT (continued)}
\\
\hline

Source & Transition & v$_{\mathrm{cen}}$ & v$_{\mathrm{fwhm}}$ & Peak T$_{\mathrm{MB}}$ & $\int T_{\mathrm{MB}} dv$ & N$_u$  \\
        &            & (km s$^{-1}$)      & (km s$^{-1}$)       & (K)     &  (K km s$^{-1}$)    &   ( 10$^{13}$ cm$^{-2}$)  \\
\hline
\hline
\endhead

M3.35+0.43 & (8,8) & 31.1 $\pm$ 0.22 & 25.33 $\pm$ 0.216 & 0.08 $\pm$ 0.001 & 2.10 $\pm$ 0.075 &  1.38 $\pm$ 0.049 \\
& (9,9) & 31.3 $\pm$ 0.22 & 25.25 $\pm$ 0.222 & 0.11 $\pm$ 0.002 & 2.93 $\pm$ 0.109 &  1.84 $\pm$ 0.068 \\
& (10,10) & 31.7 $\pm$ 0.49 & 22.71 $\pm$ 0.486 & 0.04 $\pm$ 0.002 & 1.02 $\pm$ 0.091 &  0.61 $\pm$ 0.054 \\
& (11,11) & 29.3 $\pm$ 1.34 & 29.63 $\pm$ 1.342 & 0.03 $\pm$ 0.003 & 0.87 $\pm$ 0.176 &  0.49 $\pm$ 0.099 \\

M3.14+0.41 & (8,8) & 91.0 $\pm$ 1.88 & 34.39 $\pm$ 1.877 & 0.08 $\pm$ 0.009 & 3.08 $\pm$ 0.759 &  2.02 $\pm$ 0.499 \\
& (9,9) & 91.2 $\pm$ 0.97 & 38.73 $\pm$ 0.967 & 0.12 $\pm$ 0.006 & 5.14 $\pm$ 0.604 &  3.22 $\pm$ 0.379 \\
& (10,10) & 89.9 $\pm$ 2.24 & 32.57 $\pm$ 2.244 & 0.05 $\pm$ 0.007 & 1.65 $\pm$ 0.517 &  0.98 $\pm$ 0.308 \\
& (11,11) & 90.4 $\pm$ 2.54 & 20.08 $\pm$ 2.538 & 0.03 $\pm$ 0.008 & 0.69 $\pm$ 0.357 &  0.39 $\pm$ 0.202 \\

M3.09+0.16 & (8,8) & 150.9 $\pm$ 0.21 & 24.76 $\pm$ 0.210 & 0.08 $\pm$ 0.001 & 2.22 $\pm$ 0.079 &  1.46 $\pm$ 0.052 \\
& (9,9) & 151.7 $\pm$ 0.21 & 28.19 $\pm$ 0.215 & 0.12 $\pm$ 0.002 & 3.75 $\pm$ 0.124 &  2.35 $\pm$ 0.078 \\
& (10,10) & 148.9 $\pm$ 1.11 & 21.27 $\pm$ 1.113 & 0.03 $\pm$ 0.003 & 0.71 $\pm$ 0.153 &  0.42 $\pm$ 0.091 \\
& (11,11) & 148.3 $\pm$ 1.55 & 63.17 $\pm$ 1.550 & 0.05 $\pm$ 0.003 & 3.48 $\pm$ 0.482 &  1.97 $\pm$ 0.272 \\

M1.57-0.3 & (8,8) & -31.8 $\pm$ 0.27 & 21.96 $\pm$ 0.273 & 0.09 $\pm$ 0.002 & 2.16 $\pm$ 0.109 &  1.42 $\pm$ 0.072 \\
& (9,9) & -31.0 $\pm$ 0.23 & 23.78 $\pm$ 0.225 & 0.11 $\pm$ 0.002 & 2.85 $\pm$ 0.113 &  1.79 $\pm$ 0.071 \\
& (10,10) & -33.5 $\pm$ 0.82 & 17.51 $\pm$ 0.820 & 0.04 $\pm$ 0.004 & 0.78 $\pm$ 0.145 &  0.47 $\pm$ 0.086 \\
& (11,11) & -33.2 $\pm$ 1.17 & 22.87 $\pm$ 1.166 & 0.02 $\pm$ 0.002 & 0.43 $\pm$ 0.093 &  0.24 $\pm$ 0.052 \\
& (12,12) & -28.9 $\pm$ 0.72 & 18.28 $\pm$ 0.722 & 0.04 $\pm$ 0.003 & 0.73 $\pm$ 0.116 &  0.39 $\pm$ 0.062 \\

M1.36+0.11(a) & (8,8) & 64.2 $\pm$ 1.37 & 28.80 $\pm$ 1.367 & 0.04 $\pm$ 0.004 & 1.35 $\pm$ 0.276 &  0.88 $\pm$ 0.181 \\
& (9,9) & 66.3 $\pm$ 1.32 & 35.41 $\pm$ 1.340 & 0.07 $\pm$ 0.005 & 2.62 $\pm$ 0.450 &  1.64 $\pm$ 0.282 \\
& (10,10) & 68.3 $\pm$ 3.62 & 34.20 $\pm$ 3.654 & 0.02 $\pm$ 0.003 & 0.57 $\pm$ 0.276 &  0.34 $\pm$ 0.165 \\
& (11,11) & 70.1 $\pm$ 1.75 & 31.31 $\pm$ 1.764 & 0.01 $\pm$ 0.002 & 0.49 $\pm$ 0.125 &  0.28 $\pm$ 0.070 \\
& (12,12) & 71.0 $\pm$ 1.66 & 52.75 $\pm$ 1.852 & 0.02 $\pm$ 0.001 & 1.06 $\pm$ 0.171 &  0.57 $\pm$ 0.091 \\

M1.36+0.11(b) & (8,8) & 118.1 $\pm$ 0.23 & 13.88 $\pm$ 0.234 & 0.18 $\pm$ 0.006 & 2.63 $\pm$ 0.166 &  1.73 $\pm$ 0.109 \\
& (9,9) & 119.0 $\pm$ 0.28 & 16.69 $\pm$ 0.281 & 0.23 $\pm$ 0.008 & 4.02 $\pm$ 0.261 &  2.52 $\pm$ 0.164 \\
& (10,10) & 118.0 $\pm$ 0.49 & 13.83 $\pm$ 0.490 & 0.07 $\pm$ 0.005 & 1.08 $\pm$ 0.145 &  0.64 $\pm$ 0.086 \\
& (11,11) & 120.6 $\pm$ 0.41 & 17.88 $\pm$ 0.408 & 0.05 $\pm$ 0.002 & 0.91 $\pm$ 0.083 &  0.52 $\pm$ 0.047 \\
& (12,12) & 118.8 $\pm$ 0.28 & 15.99 $\pm$ 0.296 & 0.06 $\pm$ 0.002 & 0.99 $\pm$ 0.068 &  0.53 $\pm$ 0.037 \\

M1.32-0.13(a) & (8,8) & 31.5 $\pm$ 0.28 & 12.65 $\pm$ 0.281 & 0.06 $\pm$ 0.003 & 0.77 $\pm$ 0.062 &  0.51 $\pm$ 0.041 \\
& (9,9) & 31.8 $\pm$ 0.65 & 14.71 $\pm$ 0.660 & 0.08 $\pm$ 0.007 & 1.24 $\pm$ 0.205 &  0.78 $\pm$ 0.129 \\
& (10,10) & 32.0 $\pm$ 1.12 & 12.35 $\pm$ 1.135 & 0.02 $\pm$ 0.003 & 0.24 $\pm$ 0.082 &  0.15 $\pm$ 0.049 \\
& (11,11) & 35.2 $\pm$ 1.70 & 7.03 $\pm$ 1.699 & 0.01 $\pm$ 0.003 & 0.05 $\pm$ 0.041 &  0.03 $\pm$ 0.023 \\

M1.32-0.13(b) & (8,8) & 74.6 $\pm$ 0.31 & 33.76 $\pm$ 0.318 & 0.08 $\pm$ 0.002 & 3.03 $\pm$ 0.125 &  1.99 $\pm$ 0.041 \\
& (9,9) & 74.4 $\pm$ 0.66 & 36.98 $\pm$ 0.702 & 0.12 $\pm$ 0.004 & 4.88 $\pm$ 0.402 &  3.06 $\pm$ 0.129 \\
& (10,10) & 73.0 $\pm$ 1.08 & 34.75 $\pm$ 1.130 & 0.03 $\pm$ 0.002 & 1.20 $\pm$ 0.172 &  0.71 $\pm$ 0.049 \\
& (11,11) & 78.3 $\pm$ 1.26 & 27.86 $\pm$ 1.258 & 0.02 $\pm$ 0.002 & 0.52 $\pm$ 0.103 &  0.29 $\pm$ 0.023 \\

M1.01+0.02 & (8,8) & 77.3 $\pm$ 0.12 & 18.09 $\pm$ 0.120 & 0.17 $\pm$ 0.002 & 3.22 $\pm$ 0.083 &  2.11 $\pm$ 0.055 \\
& (9,9) & 77.4 $\pm$ 0.09 & 19.41 $\pm$ 0.089 & 0.19 $\pm$ 0.002 & 3.87 $\pm$ 0.071 &  2.43 $\pm$ 0.045 \\
& (10,10) & 74.9 $\pm$ 0.35 & 19.12 $\pm$ 0.347 & 0.06 $\pm$ 0.002 & 1.13 $\pm$ 0.083 &  0.68 $\pm$ 0.049 \\
& (11,11) & 78.6 $\pm$ 0.57 & 19.68 $\pm$ 0.574 & 0.04 $\pm$ 0.003 & 0.90 $\pm$ 0.107 &  0.51 $\pm$ 0.061 \\
& (12,12) & 77.2 $\pm$ 0.43 & 20.30 $\pm$ 0.425 & 0.05 $\pm$ 0.002 & 1.12 $\pm$ 0.098 &  0.60 $\pm$ 0.052 \\

M0.89+0.14 & (8,8) & 76.5 $\pm$ 0.23 & 27.20 $\pm$ 0.230 & 0.10 $\pm$ 0.002 & 2.89 $\pm$ 0.104 &  1.90 $\pm$ 0.068 \\
& (9,9) & 76.9 $\pm$ 0.17 & 28.65 $\pm$ 0.167 & 0.13 $\pm$ 0.002 & 3.90 $\pm$ 0.099 &  2.44 $\pm$ 0.062 \\
& (10,10) & 74.5 $\pm$ 0.85 & 26.51 $\pm$ 0.850 & 0.04 $\pm$ 0.002 & 1.00 $\pm$ 0.138 &  0.60 $\pm$ 0.082 \\
& (11,11) & 77.0 $\pm$ 0.95 & 35.69 $\pm$ 0.948 & 0.03 $\pm$ 0.002 & 1.12 $\pm$ 0.141 &  0.64 $\pm$ 0.080 \\
& (12,12) & 73.9 $\pm$ 0.36 & 25.39 $\pm$ 0.361 & 0.07 $\pm$ 0.002 & 1.88 $\pm$ 0.117 &  1.01 $\pm$ 0.062 \\

M0.64-0.03 & (8,8) & 51.7 $\pm$ 0.11 & 35.44 $\pm$ 0.107 & 0.19 $\pm$ 0.001 & 7.24 $\pm$ 0.099 &  4.76 $\pm$ 0.065 \\
& (9,9) & 50.2 $\pm$ 0.14 & 34.09 $\pm$ 0.143 & 0.26 $\pm$ 0.002 & 9.38 $\pm$ 0.179 &  5.88 $\pm$ 0.112 \\
& (10,10) & 50.5 $\pm$ 0.16 & 32.09 $\pm$ 0.164 & 0.10 $\pm$ 0.001 & 3.51 $\pm$ 0.081 &  2.09 $\pm$ 0.048 \\
& (11,11) & 52.3 $\pm$ 0.45 & 33.46 $\pm$ 0.448 & 0.09 $\pm$ 0.003 & 3.28 $\pm$ 0.203 &  1.85 $\pm$ 0.115 \\
& (12,12) & 51.7 $\pm$ 0.19 & 35.14 $\pm$ 0.186 & 0.07 $\pm$ 0.001 & 2.79 $\pm$ 0.070 &  1.49 $\pm$ 0.038 \\
& (13,13) & 50.1 $\pm$ 1.41 & 32.52 $\pm$ 1.406 & 0.03 $\pm$ 0.003 & 1.13 $\pm$ 0.230 &  0.57 $\pm$ 0.116 \\

M0.48-0.01 & (8,8) & 31.2 $\pm$ 0.08 & 22.07 $\pm$ 0.079 & 0.18 $\pm$ 0.001 & 4.31 $\pm$ 0.063 &  2.83 $\pm$ 0.042 \\
& (9,9) & 30.1 $\pm$ 0.14 & 23.19 $\pm$ 0.143 & 0.24 $\pm$ 0.003 & 5.94 $\pm$ 0.152 &  3.72 $\pm$ 0.095 \\
& (10,10) & 32.1 $\pm$ 0.61 & 25.55 $\pm$ 0.614 & 0.06 $\pm$ 0.003 & 1.76 $\pm$ 0.181 &  1.05 $\pm$ 0.108 \\
& (11,11) & 27.6 $\pm$ 0.52 & 25.52 $\pm$ 0.525 & 0.07 $\pm$ 0.003 & 1.96 $\pm$ 0.174 &  1.11 $\pm$ 0.099 \\
& (12,12) & 31.1 $\pm$ 0.22 & 22.65 $\pm$ 0.217 & 0.06 $\pm$ 0.001 & 1.41 $\pm$ 0.057 &  0.75 $\pm$ 0.031 \\
& (13,13) & 32.3 $\pm$ 0.35 & 22.47 $\pm$ 0.352 & 0.02 $\pm$ 0.001 & 0.50 $\pm$ 0.034 &  0.25 $\pm$ 0.017 \\

M0.25+0.01(a) & (8,8) & 3.6 $\pm$ 0.42 & 22.40 $\pm$ 0.431 & 0.14 $\pm$ 0.003 & 3.25 $\pm$ 0.198 &  2.14 $\pm$ 0.130 \\
& (9,9) & 4.0 $\pm$ 0.27 & 25.35 $\pm$ 0.278 & 0.17 $\pm$ 0.002 & 4.61 $\pm$ 0.161 &  2.89 $\pm$ 0.101 \\
& (10,10) & 2.9 $\pm$ 0.85 & 23.32 $\pm$ 0.887 & 0.07 $\pm$ 0.004 & 1.72 $\pm$ 0.214 &  1.02 $\pm$ 0.127 \\
& (11,11) & 1.9 $\pm$ 1.20 & 39.79 $\pm$ 1.024 & 0.05 $\pm$ 0.001 & 2.12 $\pm$ 0.165 &  1.20 $\pm$ 0.093 \\
& (12,12) & 9.1 $\pm$ 0.44 & 45.23 $\pm$ 0.328 & 0.04 $\pm$ 0.000 & 1.77 $\pm$ 0.048 &  0.94 $\pm$ 0.026 \\
& (13,13) & -2.4 $\pm$ 0.44 & 10.78 $\pm$ 0.435 & 0.01 $\pm$ 0.001 & 0.16 $\pm$ 0.024 &  0.08 $\pm$ 0.012 \\

M0.25+0.01(b) & (8,8) & 29.1 $\pm$ 0.30 & 17.59 $\pm$ 0.278 & 0.17 $\pm$ 0.004 & 3.19 $\pm$ 0.162 &  2.10 $\pm$ 0.106 \\
& (9,9) & 29.5 $\pm$ 0.16 & 16.23 $\pm$ 0.146 & 0.22 $\pm$ 0.003 & 3.80 $\pm$ 0.113 &  2.38 $\pm$ 0.071 \\
& (10,10) & 29.4 $\pm$ 0.65 & 17.51 $\pm$ 0.621 & 0.08 $\pm$ 0.004 & 1.45 $\pm$ 0.169 &  0.86 $\pm$ 0.101 \\
& (11,11) & 28.9 $\pm$ 0.57 & 17.64 $\pm$ 0.696 & 0.04 $\pm$ 0.004 & 0.68 $\pm$ 0.101 &  0.38 $\pm$ 0.057 \\
& (12,12) & 32.1 $\pm$ 0.18 & 13.70 $\pm$ 0.266 & 0.02 $\pm$ 0.001 & 0.35 $\pm$ 0.024 &  0.19 $\pm$ 0.013 \\
& (13,13) & 28.0 $\pm$ 0.42 & 13.44 $\pm$ 0.423 & 0.02 $\pm$ 0.001 & 0.23 $\pm$ 0.028 &  0.12 $\pm$ 0.014 \\

M0.11-0.08 & (8,8) & 51.0 $\pm$ 0.35 & 20.89 $\pm$ 0.352 & 0.48 $\pm$ 0.017 & 10.73 $\pm$ 0.593 &  7.05 $\pm$ 0.390 \\
& (9,9) & 51.4 $\pm$ 0.75 & 21.30 $\pm$ 0.750 & 0.75 $\pm$ 0.054 & 16.94 $\pm$ 1.965 &  10.61 $\pm$ 1.231 \\
& (12,12) & 49.4 $\pm$ 0.05 & 27.51 $\pm$ 0.046 & 0.12 $\pm$ 0.000 & 3.60 $\pm$ 0.027 &  1.92 $\pm$ 0.014 \\
& (13,13) & 50.5 $\pm$ 0.09 & 28.09 $\pm$ 0.090 & 0.04 $\pm$ 0.000 & 1.34 $\pm$ 0.020 &  0.68 $\pm$ 0.010 \\

M0.07-0.08 & (8,8) & 46.6 $\pm$ 0.29 & 24.49 $\pm$ 0.285 & 0.24 $\pm$ 0.006 & 6.25 $\pm$ 0.303 &  4.11 $\pm$ 0.199 \\
& (9,9) & 47.0 $\pm$ 0.07 & 24.90 $\pm$ 0.066 & 0.35 $\pm$ 0.002 & 9.14 $\pm$ 0.102 &  5.73 $\pm$ 0.064 \\
& (10,10) & 46.6 $\pm$ 0.14 & 21.73 $\pm$ 0.142 & 0.12 $\pm$ 0.002 & 2.72 $\pm$ 0.073 &  1.62 $\pm$ 0.044 \\
& (11,11) & 47.4 $\pm$ 0.45 & 26.13 $\pm$ 0.450 & 0.10 $\pm$ 0.004 & 2.89 $\pm$ 0.216 &  1.63 $\pm$ 0.122 \\
& (12,12) & 47.5 $\pm$ 0.22 & 28.38 $\pm$ 0.215 & 0.09 $\pm$ 0.001 & 2.61 $\pm$ 0.089 &  1.39 $\pm$ 0.047 \\
& (13,13) & 44.9 $\pm$ 0.31 & 34.27 $\pm$ 0.310 & 0.04 $\pm$ 0.001 & 1.63 $\pm$ 0.071 &  0.82 $\pm$ 0.036 \\

M0.02+0.04 & (8,8) & 83.1 $\pm$ 0.11 & 23.57 $\pm$ 0.107 & 0.23 $\pm$ 0.002 & 5.68 $\pm$ 0.107 &  3.74 $\pm$ 0.070 \\
& (9,9) & 83.3 $\pm$ 0.09 & 24.83 $\pm$ 0.086 & 0.29 $\pm$ 0.002 & 7.67 $\pm$ 0.112 &  4.81 $\pm$ 0.070 \\
& (10,10) & 81.9 $\pm$ 0.78 & 25.75 $\pm$ 0.780 & 0.08 $\pm$ 0.005 & 2.14 $\pm$ 0.278 &  1.28 $\pm$ 0.166 \\
& (11,11) & 83.0 $\pm$ 0.49 & 26.75 $\pm$ 0.490 & 0.06 $\pm$ 0.002 & 1.70 $\pm$ 0.136 &  0.96 $\pm$ 0.077 \\
& (12,12) & 83.6 $\pm$ 1.08 & 29.63 $\pm$ 1.076 & 0.12 $\pm$ 0.009 & 3.71 $\pm$ 0.613 &  1.98 $\pm$ 0.327 \\
& (13,13) & 83.1 $\pm$ 0.92 & 22.19 $\pm$ 0.925 & 0.04 $\pm$ 0.003 & 0.89 $\pm$ 0.159 &  0.45 $\pm$ 0.080 \\

M359.7+0.64 & (8,8) & 2.7 $\pm$ 0.20 & 29.07 $\pm$ 0.195 & 0.13 $\pm$ 0.002 & 4.11 $\pm$ 0.120 &  2.70 $\pm$ 0.079 \\
& (9,9) & 3.5 $\pm$ 0.18 & 30.59 $\pm$ 0.182 & 0.18 $\pm$ 0.002 & 6.01 $\pm$ 0.158 &  3.77 $\pm$ 0.099 \\
& (10,10) & 3.0 $\pm$ 0.97 & 26.02 $\pm$ 0.967 & 0.06 $\pm$ 0.004 & 1.60 $\pm$ 0.255 &  0.95 $\pm$ 0.152 \\
& (11,11) & 4.5 $\pm$ 0.92 & 30.82 $\pm$ 0.925 & 0.04 $\pm$ 0.002 & 1.19 $\pm$ 0.162 &  0.67 $\pm$ 0.091 \\
& (12,12) & 3.9 $\pm$ 1.14 & 31.62 $\pm$ 1.136 & 0.07 $\pm$ 0.005 & 2.42 $\pm$ 0.402 &  1.30 $\pm$ 0.215 \\
& (13,13) & 3.9 $\pm$ 0.89 & 33.51 $\pm$ 0.889 & 0.03 $\pm$ 0.001 & 0.94 $\pm$ 0.119 &  0.47 $\pm$ 0.060 \\

M359.6-0.22(a) & (8,8) & -80.5 $\pm$ 0.24 & 27.61 $\pm$ 0.203 & 0.18 $\pm$ 0.002 & 5.21 $\pm$ 0.129 &  3.43 $\pm$ 0.085 \\
& (9,9) & -79.6 $\pm$ 0.26 & 27.88 $\pm$ 0.214 & 0.25 $\pm$ 0.003 & 7.36 $\pm$ 0.191 &  4.61 $\pm$ 0.120 \\
& (10,10) & -80.8 $\pm$ 2.84 & 27.31 $\pm$ 2.411 & 0.07 $\pm$ 0.010 & 2.02 $\pm$ 0.616 &  1.21 $\pm$ 0.367 \\
& (12,12) & -82.4 $\pm$ 0.24 & 26.19 $\pm$ 0.211 & 0.15 $\pm$ 0.002 & 4.24 $\pm$ 0.122 &  2.27 $\pm$ 0.065 \\
& (13,13) & -81.8 $\pm$ 0.58 & 25.95 $\pm$ 0.500 & 0.06 $\pm$ 0.002 & 1.64 $\pm$ 0.115 &  0.83 $\pm$ 0.058 \\
& (14,14) & -76.5 $\pm$ 0.97 & 34.45 $\pm$ 0.929 & 0.04 $\pm$ 0.002 & 1.46 $\pm$ 0.174 &  0.69 $\pm$ 0.082 \\

M359.6-0.22(b) & (8,8) & -42.3 $\pm$ 0.26 & 36.36 $\pm$ 0.248 & 0.19 $\pm$ 0.002 & 7.42 $\pm$ 0.163 &  4.88 $\pm$ 0.107 \\
& (9,9) & -41.9 $\pm$ 0.28 & 37.33 $\pm$ 0.255 & 0.28 $\pm$ 0.002 & 11.26 $\pm$ 0.245 &  7.06 $\pm$ 0.154 \\
& (10,10) & -42.7 $\pm$ 2.89 & 37.92 $\pm$ 2.727 & 0.09 $\pm$ 0.007 & 3.48 $\pm$ 0.814 &  2.07 $\pm$ 0.485 \\
& (11,11) & -76.6 $\pm$ 0.52 & 31.86 $\pm$ 0.486 & 0.05 $\pm$ 0.001 & 1.67 $\pm$ 0.108 &  0.94 $\pm$ 0.061 \\
& (11,11) & -40.0 $\pm$ 0.00 & 33.19 $\pm$ 0.420 & 0.06 $\pm$ 0.001 & 2.28 $\pm$ 0.115 &  1.29 $\pm$ 0.065 \\
& (12,12) & -43.8 $\pm$ 0.25 & 38.86 $\pm$ 0.248 & 0.19 $\pm$ 0.001 & 7.81 $\pm$ 0.170 &  4.17 $\pm$ 0.091 \\
& (13,13) & -43.5 $\pm$ 0.58 & 39.71 $\pm$ 0.560 & 0.08 $\pm$ 0.001 & 3.40 $\pm$ 0.163 &  1.71 $\pm$ 0.082 \\
& (14,14) & -40.0 $\pm$ 0.00 & 32.49 $\pm$ 0.930 & 0.04 $\pm$ 0.002 & 1.41 $\pm$ 0.170 &  0.67 $\pm$ 0.080 \\

\hline\hline
\label{tab:gbt_line_fits}
\end{longtable*}

\clearpage

\begin{longtable*}[t]{l l c c c c c}
\caption{Measured Line Parameters \& Column Densities From SWAG}
\\
\hline
Source & Transition & v$_{\mathrm{cen}}$ & v$_{\mathrm{fwhm}}$ & Peak T$_{\mathrm{MB}}$ & $\int T_{\mathrm{MB}} dv$ &  N$_u$  \\
        &            & (km s$^{-1}$)      & (km s$^{-1}$)       & (K)                     &  (K km s$^{-1}$)            & ( 10$^{13}$ cm$^{-2}$)  \\
\hline
\hline
\endfirsthead
\\
\caption{Measured Line Parameters \& Column Densities From SWAG (continued)}
\\
\hline
Source & Transition & v$_{\mathrm{cen}}$ & v$_{\mathrm{fwhm}}$ & Peak T$_{\mathrm{MB}}$ & $\int T_{\mathrm{MB}} dv$ &  N$_u$  \\
        &            & (km s$^{-1}$)      & (km s$^{-1}$)       & (K)                     &  (K km s$^{-1}$)            & ( 10$^{13}$ cm$^{-2}$)  \\
\hline
\hline
\endhead

M3.35+0.43 & (1,1) & 28.1 $\pm$ 0.59 & 35.41 $\pm$ 0.588 & 0.55 $\pm$ 0.019 & 20.62 $\pm$ 1.842 &  26.98 $\pm$ 2.410 \\
& (2,2) & 26.4 $\pm$ 0.56 & 30.25 $\pm$ 0.560 & 0.50 $\pm$ 0.019 & 16.04 $\pm$ 1.519 &  15.72 $\pm$ 1.489 \\
& (3,3) & 28.0 $\pm$ 0.22 & 25.16 $\pm$ 0.223 & 1.32 $\pm$ 0.024 & 35.24 $\pm$ 1.511 &  30.51 $\pm$ 1.309 \\
& (6,6) & 28.6 $\pm$ 1.64 & 23.56 $\pm$ 1.642 & 0.15 $\pm$ 0.021 & 3.72 $\pm$ 1.229 &  0.00 $\pm$ 0.000 \\

M3.14+0.41(a) & (1,1) & 9.6 $\pm$ 0.98 & 28.63 $\pm$ 0.980 & 0.32 $\pm$ 0.022 & 9.69 $\pm$ 1.666 &  12.67 $\pm$ 2.180 \\
& (2,2) & 9.7 $\pm$ 0.87 & 16.78 $\pm$ 0.869 & 0.24 $\pm$ 0.025 & 4.23 $\pm$ 0.950 &  4.15 $\pm$ 0.931 \\
& (3,3) & 11.7 $\pm$ 0.96 & 28.13 $\pm$ 0.956 & 0.63 $\pm$ 0.044 & 18.90 $\pm$ 3.209 &  16.36 $\pm$ 2.779 \\
& (6,6) & 12.4 $\pm$ 2.02 & 14.22 $\pm$ 2.020 & 0.07 $\pm$ 0.021 & 1.08 $\pm$ 0.643 &  0.78 $\pm$ 0.464 \\

M3.14+0.41(b) & (1,1) & 90.7 $\pm$ 0.58 & 36.68 $\pm$ 0.581 & 0.61 $\pm$ 0.020 & 23.71 $\pm$ 2.043 &  12.67 $\pm$ 2.180 \\
& (2,2) & 90.6 $\pm$ 0.55 & 32.35 $\pm$ 0.553 & 0.52 $\pm$ 0.018 & 17.77 $\pm$ 1.587 &  4.15 $\pm$ 0.931 \\
& (3,3) & 90.4 $\pm$ 0.43 & 37.55 $\pm$ 0.431 & 1.62 $\pm$ 0.038 & 64.61 $\pm$ 4.071 &  16.36 $\pm$ 2.779 \\
& (6,6) & 90.5 $\pm$ 0.74 & 22.80 $\pm$ 0.743 & 0.25 $\pm$ 0.016 & 5.98 $\pm$ 0.917 &  4.32 $\pm$ 0.662 \\

M3.09+0.16 & (1,1) & 151.5 $\pm$ 0.99 & 41.52 $\pm$ 0.994 & 0.96 $\pm$ 0.047 & 42.61 $\pm$ 5.774 &  55.75 $\pm$ 7.554 \\
& (2,2) & 151.3 $\pm$ 0.97 & 40.83 $\pm$ 0.973 & 0.68 $\pm$ 0.033 & 29.38 $\pm$ 3.941 &  28.79 $\pm$ 3.862 \\
& (3,3) & 152.2 $\pm$ 0.72 & 35.16 $\pm$ 0.725 & 2.03 $\pm$ 0.085 & 76.14 $\pm$ 8.420 &  65.92 $\pm$ 7.290 \\
& (6,6) & 153.5 $\pm$ 0.87 & 26.70 $\pm$ 0.874 & 0.34 $\pm$ 0.023 & 9.60 $\pm$ 1.546 &  0.00 $\pm$ 0.000 \\

M1.57-0.3 & (1,1) & -27.6 $\pm$ 1.78 & 35.99 $\pm$ 1.776 & 1.41 $\pm$ 0.142 & 54.08 $\pm$ 9.942 &  70.76 $\pm$ 13.007 \\
& (2,2) & -28.0 $\pm$ 1.53 & 30.16 $\pm$ 1.528 & 1.13 $\pm$ 0.117 & 36.44 $\pm$ 6.707 &  35.71 $\pm$ 6.573 \\
& (3,3) & -27.6 $\pm$ 1.23 & 24.81 $\pm$ 1.229 & 3.30 $\pm$ 0.333 & 87.23 $\pm$ 15.319 &  75.53 $\pm$ 13.263 \\
& (4,4) & -28.7 $\pm$ 1.07 & 20.34 $\pm$ 1.072 & 0.56 $\pm$ 0.060 & 12.11 $\pm$ 2.217 &  9.72 $\pm$ 1.779 \\
& (5,5) & -28.8 $\pm$ 1.02 & 20.64 $\pm$ 1.019 & 0.42 $\pm$ 0.042 & 9.13 $\pm$ 1.567 &  6.92 $\pm$ 1.188 \\
& (6,6) & -28.2 $\pm$ 1.41 & 21.04 $\pm$ 1.410 & 0.60 $\pm$ 0.082 & 13.42 $\pm$ 3.132 &  9.68 $\pm$ 2.261 \\

M1.36+0.11(a) & (1,1) & 65.8 $\pm$ 2.99 & 32.20 $\pm$ 2.866 & 0.78 $\pm$ 0.103 & 26.75 $\pm$ 7.378 &  35.00 $\pm$ 9.653 \\
& (2,2) & 62.7 $\pm$ 1.59 & 20.31 $\pm$ 1.635 & 0.84 $\pm$ 0.126 & 18.15 $\pm$ 4.800 &  17.79 $\pm$ 4.705 \\
& (3,3) & 70.6 $\pm$ 2.55 & 41.12 $\pm$ 2.819 & 1.36 $\pm$ 0.151 & 59.61 $\pm$ 13.534 &  51.61 $\pm$ 11.718 \\
& (4,4) & 88.2 $\pm$ 4.52 & 59.92 $\pm$ 3.884 & 0.23 $\pm$ 0.023 & 14.68 $\pm$ 3.264 &  11.79 $\pm$ 2.620 \\
& (5,5) & 84.0 $\pm$ 3.62 & 51.35 $\pm$ 3.793 & 0.17 $\pm$ 0.017 & 9.12 $\pm$ 2.107 &  6.92 $\pm$ 1.598 \\
& (6,6) & 65.2 $\pm$ 2.42 & 27.62 $\pm$ 2.507 & 0.32 $\pm$ 0.056 & 9.39 $\pm$ 2.960 &  6.78 $\pm$ 2.136 \\

M1.36+0.11(b) & (1,1) & 113.4 $\pm$ 2.83 & 4.07 $\pm$ 2.936 & 0.93 $\pm$ 0.092 & 40.17 $\pm$ 8.902 &  52.55 $\pm$ 11.647 \\
& (2,2) & 111.7 $\pm$ 3.02 & 50.94 $\pm$ 3.378 & 0.73 $\pm$ 0.079 & 39.61 $\pm$ 8.981 &  38.82 $\pm$ 8.802 \\
& (3,3) & 117.6 $\pm$ 1.27 & 24.64 $\pm$ 1.271 & 2.14 $\pm$ 0.190 & 56.11 $\pm$ 9.392 &  48.58 $\pm$ 8.132 \\
& (4,4) & 119.4 $\pm$ 1.00 & 13.49 $\pm$ 1.290 & 0.32 $\pm$ 0.058 & 4.62 $\pm$ 1.395 &  3.70 $\pm$ 1.120 \\
& (5,5) & 120.2 $\pm$ 0.75 & 14.98 $\pm$ 0.921 & 0.31 $\pm$ 0.037 & 4.94 $\pm$ 0.974 &  3.74 $\pm$ 0.738 \\
& (6,6) & 119.1 $\pm$ 2.19 & 33.73 $\pm$ 2.291 & 0.39 $\pm$ 0.051 & 13.98 $\pm$ 3.371 &  10.09 $\pm$ 2.433 \\

M1.32-0.13(a) & (1,1) & 36.1 $\pm$ 9.88 & 89.78 $\pm$ 6.881 & 0.62 $\pm$ 0.057 & 59.48 $\pm$ 13.660 &  77.82 $\pm$ 17.871 \\
& (2,2) & 42.5 $\pm$ 5.60 & 94.20 $\pm$ 3.645 & 0.53 $\pm$ 0.042 & 53.64 $\pm$ 8.279 &  52.57 $\pm$ 8.114 \\
& (3,3) & 33.9 $\pm$ 0.59 & 18.34 $\pm$ 0.594 & 1.65 $\pm$ 0.103 & 32.16 $\pm$ 3.484 &  27.84 $\pm$ 3.017 \\
& (4,4) & 32.0 $\pm$ 0.62 & 15.14 $\pm$ 0.627 & 0.18 $\pm$ 0.015 & 2.93 $\pm$ 0.408 &  2.35 $\pm$ 0.328 \\
& (5,5) & 26.6 $\pm$ 4.32 & 150.00 $\pm$ 0.000 & 0.08 $\pm$ 0.004 & 12.70 $\pm$ 1.347 &  9.63 $\pm$ 1.021 \\
& (6,6) & 34.1 $\pm$ 0.87 & 12.60 $\pm$ 0.870 & 0.42 $\pm$ 0.058 & 5.60 $\pm$ 1.284 &  4.04 $\pm$ 0.927 \\

M1.32-0.13(b) & (1,1) & 84.7 $\pm$ 1.41 & 35.46 $\pm$ 2.395 & 0.97 $\pm$ 0.172 & 36.44 $\pm$ 9.083 &  47.68 $\pm$ 11.883 \\
& (2,2) & 82.5 $\pm$ 1.06 & 28.68 $\pm$ 1.459 & 0.84 $\pm$ 0.087 & 25.68 $\pm$ 4.351 &  25.16 $\pm$ 4.264 \\
& (3,3) & 79.8 $\pm$ 0.61 & 35.55 $\pm$ 0.641 & 2.19 $\pm$ 0.075 & 82.92 $\pm$ 5.317 &  71.79 $\pm$ 4.603 \\
& (4,4) & 76.7 $\pm$ 0.35 & 28.56 $\pm$ 0.355 & 0.44 $\pm$ 0.011 & 13.44 $\pm$ 0.598 &  10.79 $\pm$ 0.480 \\
& (5,5) & 77.4 $\pm$ 0.35 & 23.83 $\pm$ 0.401 & 0.31 $\pm$ 0.010 & 7.93 $\pm$ 0.447 &  6.01 $\pm$ 0.339 \\
& (6,6) & 75.7 $\pm$ 1.03 & 28.84 $\pm$ 1.049 & 0.53 $\pm$ 0.039 & 16.18 $\pm$ 2.106 &  11.67 $\pm$ 1.520 \\

M1.01+0.02(a) & (1,1) & 61.4 $\pm$ 11.68 & 34.24 $\pm$ 6.665 & 0.45 $\pm$ 0.149 & 16.25 $\pm$ 9.361 &  21.26 $\pm$ 12.247 \\
& (2,2) & 60.1 $\pm$ 2.43 & 28.61 $\pm$ 2.631 & 0.36 $\pm$ 0.045 & 10.83 $\pm$ 2.957 &  10.61 $\pm$ 2.898 \\
& (3,3) & 57.1 $\pm$ 1.02 & 22.14 $\pm$ 1.104 & 0.81 $\pm$ 0.060 & 19.01 $\pm$ 2.862 &  16.46 $\pm$ 2.478 \\
& (4,4) & 53.8 $\pm$ 0.87 & 18.71 $\pm$ 0.908 & 0.14 $\pm$ 0.010 & 2.69 $\pm$ 0.397 &  2.16 $\pm$ 0.318 \\

M1.01+0.02(b) & (1,1) & 86.8 $\pm$ 3.99 & 25.97 $\pm$ 2.161 & 0.80 $\pm$ 0.298 & 22.13 $\pm$ 9.393 &  28.95 $\pm$ 12.289 \\
& (2,2) & 84.7 $\pm$ 0.64 & 12.22 $\pm$ 0.666 & 0.75 $\pm$ 0.077 & 9.74 $\pm$ 1.682 &  9.55 $\pm$ 1.649 \\
& (3,3) & 84.1 $\pm$ 0.46 & 16.94 $\pm$ 0.461 & 1.58 $\pm$ 0.068 & 28.51 $\pm$ 2.349 &  24.69 $\pm$ 2.034 \\
& (4,4) & 81.2 $\pm$ 0.42 & 19.91 $\pm$ 0.443 & 0.29 $\pm$ 0.010 & 6.15 $\pm$ 0.415 &  4.94 $\pm$ 0.333 \\
& (5,5) & 81.5 $\pm$ 0.44 & 19.18 $\pm$ 0.436 & 0.21 $\pm$ 0.010 & 4.38 $\pm$ 0.345 &  3.32 $\pm$ 0.261 \\
& (6,6) & 77.9 $\pm$ 1.67 & 16.60 $\pm$ 1.672 & 0.28 $\pm$ 0.058 & 4.96 $\pm$ 1.704 &  3.58 $\pm$ 1.230 \\

M0.89+0.14 & (1,1) & 80.5 $\pm$ 0.74 & 32.88 $\pm$ 0.743 & 1.19 $\pm$ 0.055 & 41.49 $\pm$ 3.445 &  54.28 $\pm$ 4.507 \\
& (2,2) & 82.0 $\pm$ 1.02 & 31.18 $\pm$ 1.023 & 0.72 $\pm$ 0.048 & 23.85 $\pm$ 2.854 &  23.38 $\pm$ 2.797 \\
& (3,3) & 78.3 $\pm$ 0.38 & 29.53 $\pm$ 0.379 & 1.81 $\pm$ 0.047 & 56.82 $\pm$ 2.641 &  49.20 $\pm$ 2.286 \\
& (4,4) & 78.8 $\pm$ 0.42 & 29.17 $\pm$ 0.425 & 0.45 $\pm$ 0.013 & 13.84 $\pm$ 0.728 &  11.11 $\pm$ 0.585 \\
& (5,5) & 79.5 $\pm$ 0.53 & 25.36 $\pm$ 0.529 & 0.25 $\pm$ 0.011 & 6.88 $\pm$ 0.510 &  5.22 $\pm$ 0.387 \\
& (6,6) & 77.8 $\pm$ 1.34 & 23.94 $\pm$ 1.344 & 0.46 $\pm$ 0.053 & 11.70 $\pm$ 2.321 &  8.45 $\pm$ 1.675 \\

M0.64-0.03 & (1,1) & 53.1 $\pm$ 0.27 & 48.46 $\pm$ 0.269 & 3.60 $\pm$ 0.041 & 185.58 $\pm$ 4.029 &  242.80 $\pm$ 5.271 \\
& (2,2) & 52.1 $\pm$ 0.36 & 50.82 $\pm$ 0.358 & 2.58 $\pm$ 0.037 & 139.49 $\pm$ 3.879 &  136.71 $\pm$ 3.801 \\
& (3,3) & 53.0 $\pm$ 0.19 & 45.93 $\pm$ 0.191 & 5.44 $\pm$ 0.046 & 265.79 $\pm$ 4.275 &  230.12 $\pm$ 3.701 \\
& (4,4) & 50.6 $\pm$ 0.18 & 36.75 $\pm$ 0.175 & 1.09 $\pm$ 0.011 & 42.49 $\pm$ 0.757 &  34.10 $\pm$ 0.608 \\
& (5,5) & 51.1 $\pm$ 0.17 & 36.17 $\pm$ 0.168 & 0.79 $\pm$ 0.008 & 30.45 $\pm$ 0.528 &  23.09 $\pm$ 0.400 \\
& (6,6) & 52.1 $\pm$ 0.66 & 34.04 $\pm$ 0.659 & 0.95 $\pm$ 0.038 & 34.42 $\pm$ 2.462 &  24.84 $\pm$ 1.777 \\

M0.25+0.01(a) & (1,1) & 0.8 $\pm$ 1.60 & 33.61 $\pm$ 0.796 & 2.21 $\pm$ 0.194 & 79.23 $\pm$ 8.327 &  103.66 $\pm$ 10.894 \\
& (2,2) & -2.9 $\pm$ 1.06 & 24.40 $\pm$ 0.807 & 1.34 $\pm$ 0.141 & 34.73 $\pm$ 4.662 &  34.04 $\pm$ 4.569 \\
& (3,3) & 2.1 $\pm$ 0.17 & 21.35 $\pm$ 0.171 & 4.79 $\pm$ 0.054 & 108.84 $\pm$ 2.561 &  94.23 $\pm$ 2.217 \\
& (4,4) & 4.6 $\pm$ 0.30 & 23.01 $\pm$ 0.316 & 0.55 $\pm$ 0.011 & 13.46 $\pm$ 0.547 &  10.80 $\pm$ 0.439 \\
& (5,5) & 2.9 $\pm$ 0.25 & 21.46 $\pm$ 0.258 & 0.43 $\pm$ 0.007 & 9.89 $\pm$ 0.351 &  7.50 $\pm$ 0.266 \\
& (6,6) & 2.3 $\pm$ 1.02 & 21.23 $\pm$ 1.026 & 0.70 $\pm$ 0.046 & 15.79 $\pm$ 2.229 &  11.40 $\pm$ 1.609 \\

M0.25+0.01(b) & (1,1) & 31.9 $\pm$ 1.64 & 38.60 $\pm$ 0.880 & 2.72 $\pm$ 0.133 & 111.85 $\pm$ 8.251 &  146.34 $\pm$ 10.795 \\
& (2,2) & 26.6 $\pm$ 1.24 & 38.81 $\pm$ 0.910 & 2.19 $\pm$ 0.052 & 90.52 $\pm$ 5.679 &  88.72 $\pm$ 5.566 \\
& (3,3) & 30.3 $\pm$ 0.17 & 21.27 $\pm$ 0.169 & 4.82 $\pm$ 0.054 & 109.11 $\pm$ 2.553 &  94.47 $\pm$ 2.211 \\
& (4,4) & 31.2 $\pm$ 0.15 & 17.10 $\pm$ 0.143 & 0.96 $\pm$ 0.012 & 17.45 $\pm$ 0.436 &  14.00 $\pm$ 0.350 \\
& (5,5) & 30.8 $\pm$ 0.14 & 20.44 $\pm$ 0.143 & 0.75 $\pm$ 0.008 & 16.33 $\pm$ 0.338 &  12.38 $\pm$ 0.256 \\
& (6,6) & 29.6 $\pm$ 0.92 & 20.48 $\pm$ 0.915 & 0.76 $\pm$ 0.047 & 16.59 $\pm$ 2.166 &  11.97 $\pm$ 1.563 \\

M0.11-0.08 (a) & (1,1) & 49.8 $\pm$ 0.15 & 38.19 $\pm$ 0.145 & 5.12 $\pm$ 0.040 & 208.14 $\pm$ 2.973 &  272.31 $\pm$ 3.889 \\
& (2,2) & 49.7 $\pm$ 0.15 & 35.38 $\pm$ 0.149 & 4.14 $\pm$ 0.036 & 155.96 $\pm$ 2.444 &  152.86 $\pm$ 2.396 \\
& (3,3) & 50.3 $\pm$ 0.05 & 26.33 $\pm$ 0.050 & 11.68 $\pm$ 0.045 & 327.48 $\pm$ 2.217 &  283.53 $\pm$ 1.920 \\
& (4,4) & 49.6 $\pm$ 0.06 & 23.80 $\pm$ 0.056 & 2.17 $\pm$ 0.010 & 54.88 $\pm$ 0.452 &  44.05 $\pm$ 0.363 \\
& (5,5) & 50.3 $\pm$ 0.07 & 23.48 $\pm$ 0.068 & 1.42 $\pm$ 0.008 & 35.52 $\pm$ 0.361 &  26.93 $\pm$ 0.274 \\
& (6,6) & 49.4 $\pm$ 0.30 & 23.23 $\pm$ 0.298 & 1.99 $\pm$ 0.052 & 49.22 $\pm$ 2.220 &  35.52 $\pm$ 1.602 \\

M0.07-0.08 & (1,1) & 47.7 $\pm$ 0.15 & 36.82 $\pm$ 0.152 & 5.92 $\pm$ 0.050 & 231.95 $\pm$ 3.573 &  303.46 $\pm$ 4.675 \\
& (2,2) & 47.7 $\pm$ 0.19 & 32.57 $\pm$ 0.191 & 4.37 $\pm$ 0.052 & 151.41 $\pm$ 3.258 &  148.40 $\pm$ 3.193 \\
& (3,3) & 47.3 $\pm$ 0.08 & 26.61 $\pm$ 0.062 & 10.00 $\pm$ 0.000 & 283.27 $\pm$ 2.099 &  245.25 $\pm$ 1.818 \\
& (4,4) & 46.3 $\pm$ 0.07 & 21.91 $\pm$ 0.068 & 1.92 $\pm$ 0.012 & 44.73 $\pm$ 0.489 &  35.90 $\pm$ 0.392 \\
& (5,5) & 47.1 $\pm$ 0.08 & 23.40 $\pm$ 0.084 & 1.23 $\pm$ 0.009 & 30.73 $\pm$ 0.387 &  23.30 $\pm$ 0.293 \\
& (6,6) & 46.8 $\pm$ 0.36 & 21.35 $\pm$ 0.365 & 1.52 $\pm$ 0.053 & 34.60 $\pm$ 2.062 &  24.97 $\pm$ 1.488 \\

M0.02+0.04 & (1,1) & 85.2 $\pm$ 0.55 & 30.36 $\pm$ 0.548 & 2.57 $\pm$ 0.095 & 83.08 $\pm$ 5.453 &  108.69 $\pm$ 7.134 \\
& (2,2) & 85.0 $\pm$ 0.65 & 32.10 $\pm$ 0.647 & 2.01 $\pm$ 0.083 & 68.84 $\pm$ 5.082 &  67.46 $\pm$ 4.980 \\
& (3,3) & 84.5 $\pm$ 0.29 & 25.09 $\pm$ 0.291 & 5.62 $\pm$ 0.133 & 150.11 $\pm$ 6.172 &  129.97 $\pm$ 5.344 \\
& (4,4) & 84.7 $\pm$ 0.24 & 21.55 $\pm$ 0.244 & 0.95 $\pm$ 0.022 & 21.74 $\pm$ 0.861 &  17.45 $\pm$ 0.691 \\
& (5,5) & 83.9 $\pm$ 0.24 & 28.30 $\pm$ 0.241 & 0.71 $\pm$ 0.012 & 21.26 $\pm$ 0.652 &  16.12 $\pm$ 0.495 \\
& (6,6) & 85.1 $\pm$ 0.63 & 23.85 $\pm$ 0.625 & 0.98 $\pm$ 0.052 & 24.89 $\pm$ 2.304 &  17.96 $\pm$ 1.663 \\

M359.7+0.64 & (1,1) & 2.5 $\pm$ 1.61 & 33.46 $\pm$ 1.613 & 0.86 $\pm$ 0.084 & 30.53 $\pm$ 5.424 &  39.94 $\pm$ 7.096 \\
& (2,2) & 3.5 $\pm$ 1.84 & 36.79 $\pm$ 1.836 & 0.74 $\pm$ 0.075 & 28.84 $\pm$ 5.378 &  28.27 $\pm$ 5.271 \\
& (3,3) & 3.0 $\pm$ 1.19 & 27.24 $\pm$ 1.190 & 2.54 $\pm$ 0.226 & 73.56 $\pm$ 11.526 &  63.69 $\pm$ 9.979 \\
& (4,4) & 3.0 $\pm$ 1.39 & 25.75 $\pm$ 1.385 & 0.36 $\pm$ 0.040 & 9.94 $\pm$ 1.904 &  7.98 $\pm$ 1.529 \\
& (5,5) & 2.6 $\pm$ 1.33 & 35.03 $\pm$ 1.328 & 0.32 $\pm$ 0.025 & 12.03 $\pm$ 1.692 &  9.12 $\pm$ 1.283 \\
& (6,6) & 2.9 $\pm$ 1.76 & 25.53 $\pm$ 1.757 & 0.50 $\pm$ 0.070 & 13.48 $\pm$ 3.302 &  9.73 $\pm$ 2.383 \\

M359.6-0.22(a) & (1,1) & -85.1 $\pm$ 0.89 & 22.32 $\pm$ 1.295 & 0.99 $\pm$ 0.123 & 23.49 $\pm$ 4.546 &  30.73 $\pm$ 5.947 \\
& (2,2) & -83.8 $\pm$ 0.74 & 20.96 $\pm$ 0.980 & 1.12 $\pm$ 0.103 & 24.88 $\pm$ 3.784 &  24.38 $\pm$ 3.709 \\
& (3,3) & -83.3 $\pm$ 0.29 & 21.45 $\pm$ 0.315 & 5.06 $\pm$ 0.160 & 115.48 $\pm$ 5.725 &  99.98 $\pm$ 4.956 \\
& (4,4) & -83.1 $\pm$ 0.42 & 21.26 $\pm$ 0.399 & 0.67 $\pm$ 0.021 & 15.08 $\pm$ 0.892 &  12.11 $\pm$ 0.716 \\
& (5,5) & -84.2 $\pm$ 0.51 & 24.42 $\pm$ 0.496 & 0.53 $\pm$ 0.023 & 13.73 $\pm$ 0.938 &  10.41 $\pm$ 0.711 \\
& (6,6) & -81.9 $\pm$ 0.77 & 21.56 $\pm$ 0.753 & 1.04 $\pm$ 0.063 & 23.92 $\pm$ 2.660 &  17.26 $\pm$ 1.920 \\

M359.6-0.22(b) & (1,1) & -55.0 $\pm$ 2.87 & 67.84 $\pm$ 1.831 & 1.11 $\pm$ 0.057 & 79.97 $\pm$ 7.741 &  104.63 $\pm$ 10.127 \\
& (2,2) & -55.8 $\pm$ 3.07 & 71.48 $\pm$ 1.972 & 0.85 $\pm$ 0.055 & 64.88 $\pm$ 7.284 &  63.58 $\pm$ 7.139 \\
& (3,3) & -46.1 $\pm$ 1.12 & 48.40 $\pm$ 1.095 & 2.87 $\pm$ 0.073 & 148.00 $\pm$ 9.688 &  128.14 $\pm$ 8.388 \\
& (4,4) & -46.3 $\pm$ 0.98 & 37.40 $\pm$ 1.048 & 0.40 $\pm$ 0.015 & 15.87 $\pm$ 1.351 &  12.74 $\pm$ 1.084 \\
& (5,5) & -44.6 $\pm$ 1.26 & 47.99 $\pm$ 1.215 & 0.40 $\pm$ 0.011 & 20.30 $\pm$ 1.485 &  15.39 $\pm$ 1.126 \\
& (6,6) & -42.1 $\pm$ 1.55 & 39.61 $\pm$ 1.690 & 0.73 $\pm$ 0.045 & 30.87 $\pm$ 4.130 &  22.28 $\pm$ 2.981 \\
\hline
\hline
\label{tab:swag_line_fits}
\end{longtable*}

\clearpage

\begin{table}[t]
\caption{Measured Line Parameters and Column Densities From SWAG Data with Hyperfine Fitting--M0.48}
\begin{tabular}{l c c c c c c}
\hline
Source & Transition & v$_{\mathrm{cen}}$ & v$_{\mathrm{fwhm}}$ & T$_{ex}$ & $\tau$& N$_u$  \\
        &            & (km s$^{-1}$)      & (km s$^{-1}$)       & (K)     &     &   ( 10$^{13}$ cm$^{-2}$)  \\
\hline
\hline

M0.48-0.01 & (1,1) & 30.0 $\pm$ 0.01 & 7.64 $\pm$ 0.010 & 10.05 $\pm$ 0.014 &  14.24 $\pm$ 0.105 & 1485.92 $\pm$ 11.970 \\
& (2,2) & 30.2 $\pm$ 0.01 & 8.22 $\pm$ 0.007 & 8.66 $\pm$ 0.014 &  9.79 $\pm$ 0.061 & 709.96 $\pm$ 4.835 \\
& (3,3) & 30.1 $\pm$ 0.01 & 10.84 $\pm$ 0.009 & 30.75 $\pm$ 0.217 &  1.30 $\pm$ 0.018 & 388.67 $\pm$ 6.001 \\
& (4,4) & 29.8 $\pm$ 0.07 & 13.13 $\pm$ 0.069 & -- & -- &  19.63 $\pm$ 0.346 \\
& (5,5) & 29.6 $\pm$ 0.09 & 16.69 $\pm$ 0.091 & -- & -- &  14.24 $\pm$ 0.265 \\
& (6,6) & 36.8 $\pm$ 0.37 & 15.51 $\pm$ 0.368 & -- & -- &  16.15 $\pm$ 1.300 \\

\hline
\hline
\end{tabular}

\label{tab:swag_M0.48}
\end{table}

\clearpage

\begin{figure}[tbh]
    \centering
    \includegraphics[width=\textwidth]{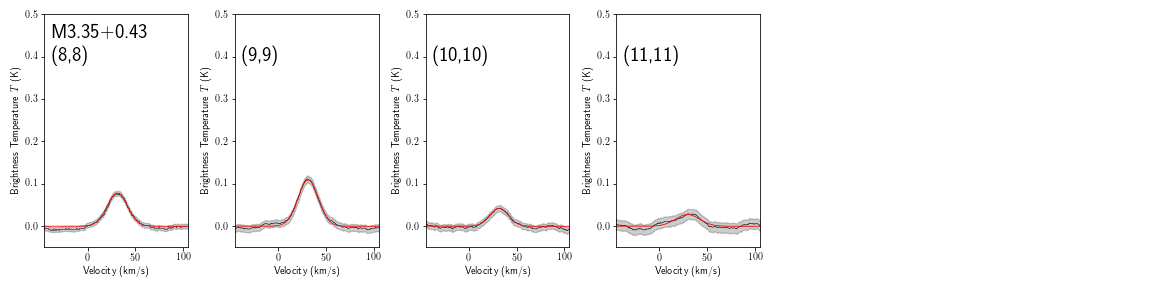}
    \caption{Source M3.35+0.43 showing (8,8)-(11,11) transitions.}
    \label{fig:fig2a}
\end{figure}

\begin{figure}[tbh]
    \centering
    \includegraphics[width=\textwidth]{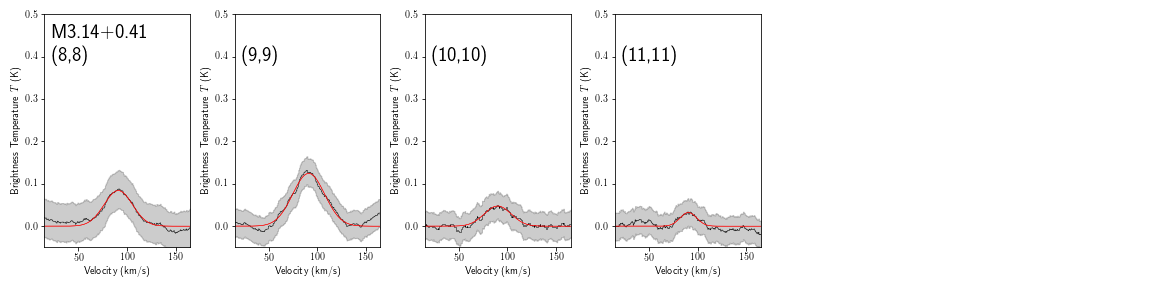}
    \caption{Source M3.14+0.41 showing (8,8)-(11,11) transitions.}
    \label{fig:fig2b}
\end{figure}

\begin{figure}[tbh]
    \centering
    \includegraphics[width=\textwidth]{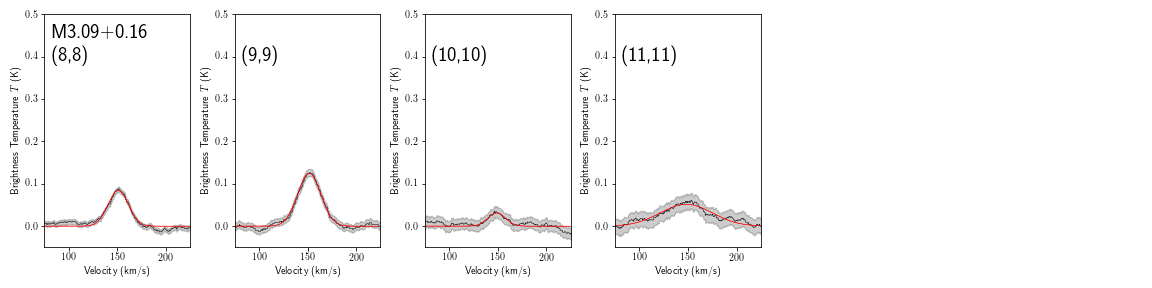}
    \caption{Source M3.09+0.16 showing (8,8)-(11,11) transitions.}
    \label{fig:fig2c}
\end{figure}

\begin{figure}[tbh]
    \centering
    \includegraphics[width=\textwidth]{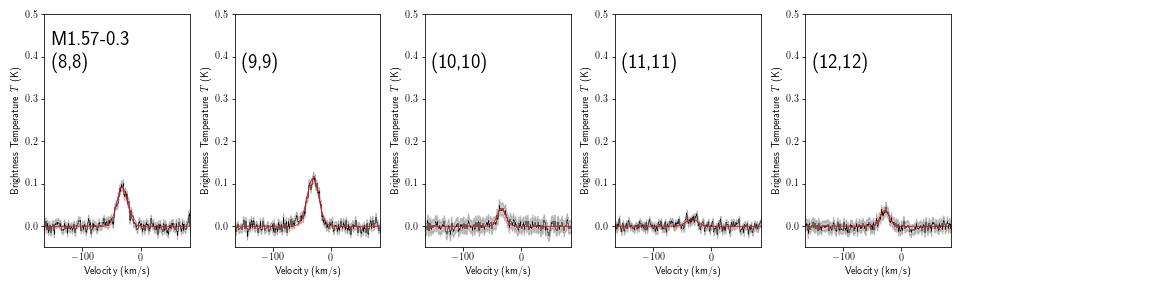}
    \caption{Source M1.57-0.3 showing (8,8)-(12,12) transitions.}
    \label{fig:fig2d}
\end{figure}

\begin{figure}[tbh]
    \centering
    \includegraphics[width=\textwidth]{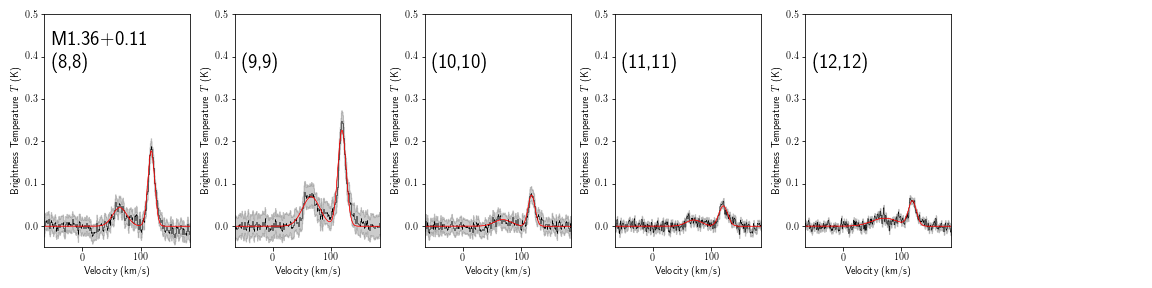}
    \caption{Source M1.36+0.11 showing (8,8)-(12,12) transitions.}
    \label{fig:fig2e}
\end{figure}

\begin{figure}[tbh]
    \centering
    \includegraphics[width=\textwidth]{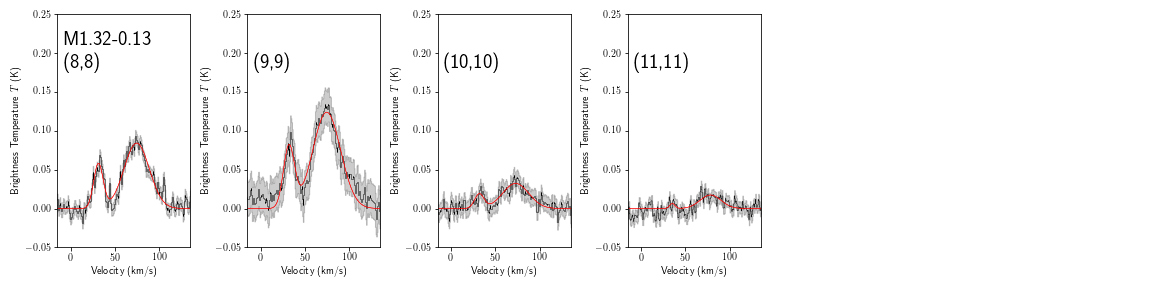}
    \caption{Source M1.32-0.13 showing (8,8)-(11,11) transitions.}
    \label{fig:fig2f}
\end{figure}

\begin{figure}[tbh]
    \centering
    \includegraphics[width=\textwidth]{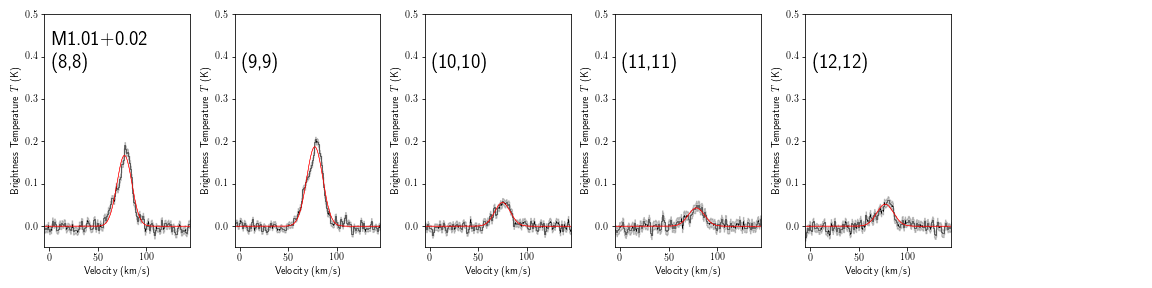}
    \caption{Source M1.01-0.13 showing (8,8)-(12,12) transitions.}
    \label{fig:fig2g}
\end{figure}

\begin{figure}[tbh]
    \centering
    \includegraphics[width=\textwidth]{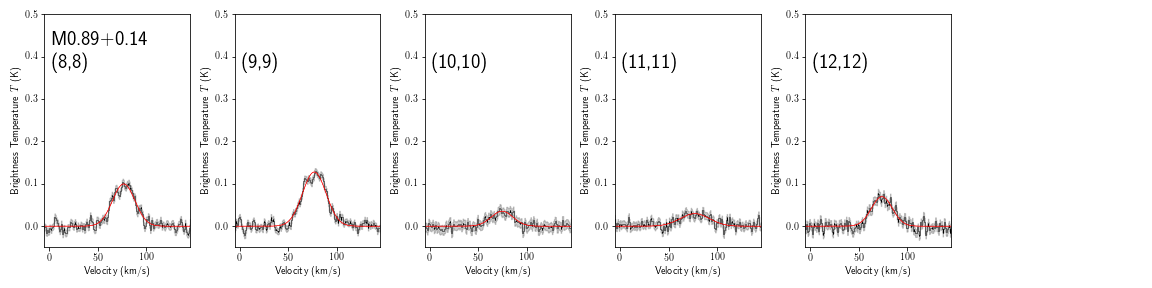}
    \caption{Source M0.89+0.14 showing (8,8)-(12,12) transitions.}
    \label{fig:fig2h}
\end{figure}

\begin{figure}[tbh]
    \centering
    \includegraphics[width=\textwidth]{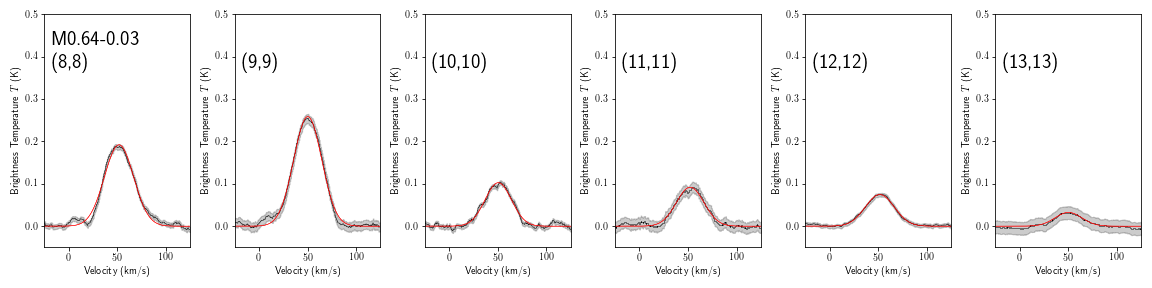}
    \caption{Source M0.64-0.03 showing (8,8)-(13,13) transitions.}
    \label{fig:fig2i}
\end{figure}

\begin{figure}[tbh]
    \centering
    \includegraphics[width=\textwidth]{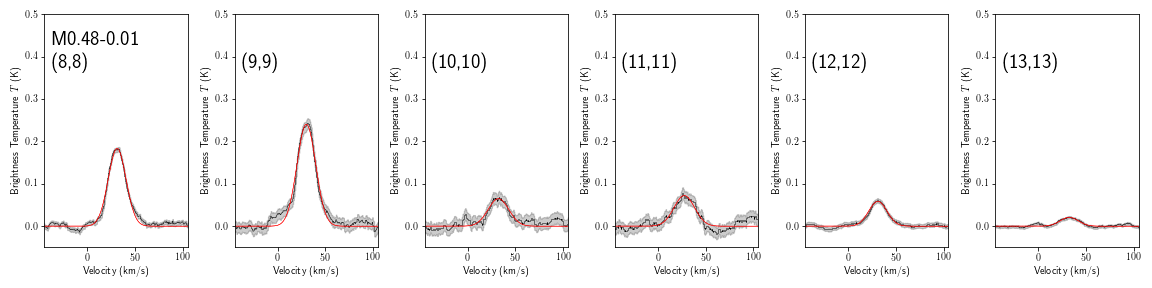}
    \caption{Source M0.48-0.01 showing (8,8)-(13,13) transitions.}
    \label{fig:fig2j}
\end{figure}

\begin{figure}[tbh]
    \centering
    \includegraphics[width=\textwidth]{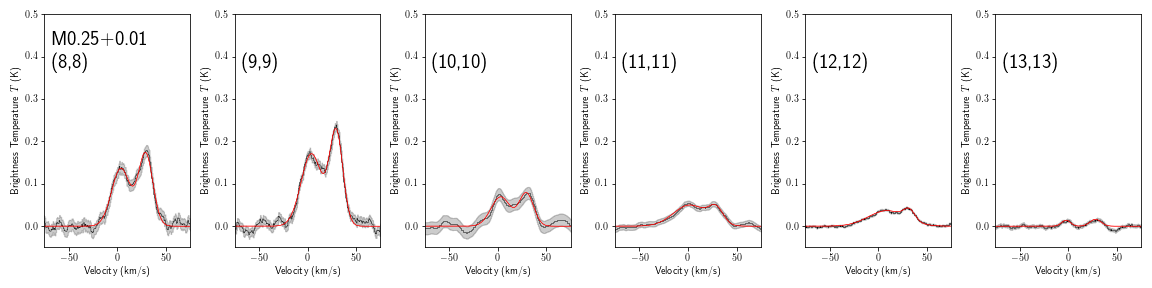}
    \caption{Source M0.25+0.01 showing (8,8)-(13,13) transitions.}
    \label{fig:fig2k}
\end{figure}

\begin{figure}[tbh]
    \centering
    \includegraphics[width=\textwidth]{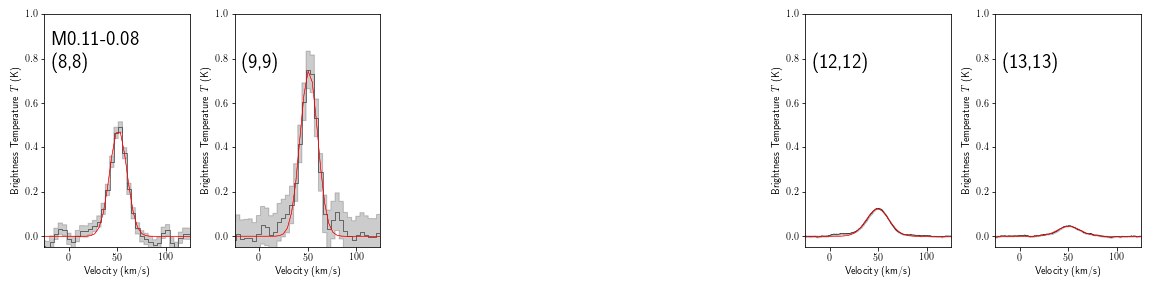}
    \caption{Source M0.11-0.08 showing (8,8), (9,9), (12,12), and (13,13) transitions.}
    \label{fig:fig2l}
\end{figure}

\begin{figure}[tbh]
    \centering
    \includegraphics[width=\textwidth]{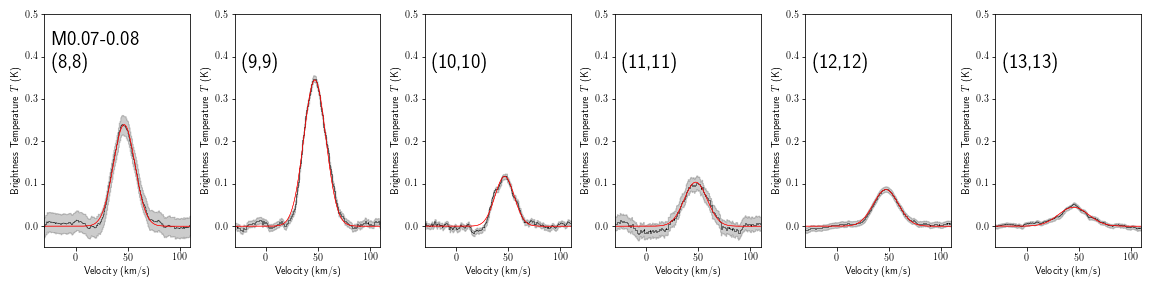}
    \caption{Source M0.07-0.08 showing (8,8)-(13,13) transitions.}
    \label{fig:fig2m}
\end{figure}

\begin{figure}[tbh]
    \centering
    \includegraphics[width=\textwidth]{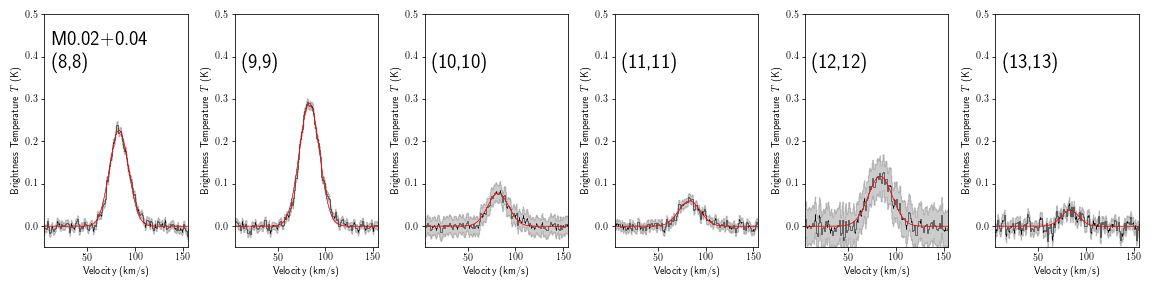}
    \caption{Source M0.02+0.04 showing (8,8)-(13,13) transitions.}
    \label{fig:fig2n}
\end{figure}

\begin{figure}[tbh]
    \centering
    \includegraphics[width=\textwidth]{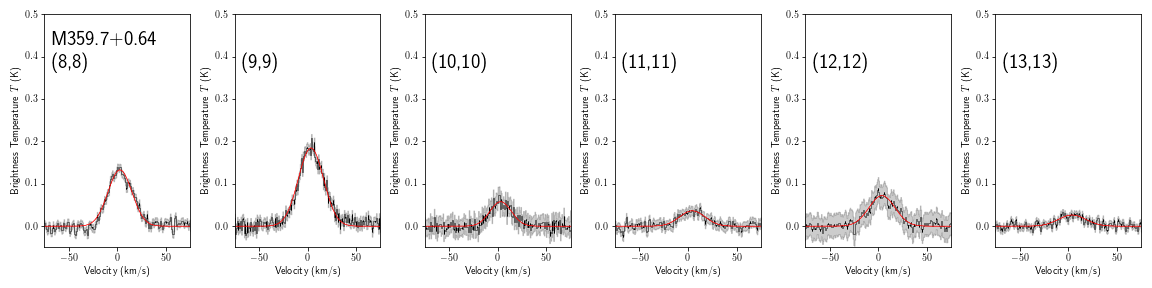}
    \caption{Source M359.7+0.64 showing (8,8)-(13,13) transitions.}
    \label{fig:fig2o}
\end{figure}

\begin{figure}[tbh]
    \centering
    \includegraphics[width=\textwidth]{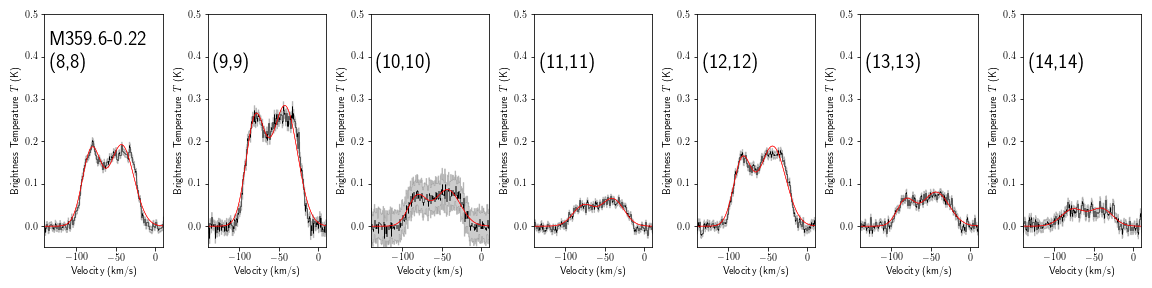}
    \caption{Source M359.6-0.22 showing (8,8)-(14,14) transitions.}
    \label{fig:fig2p}
\end{figure}

\begin{figure}[tbh]
    \centering
    \includegraphics[width=\textwidth]{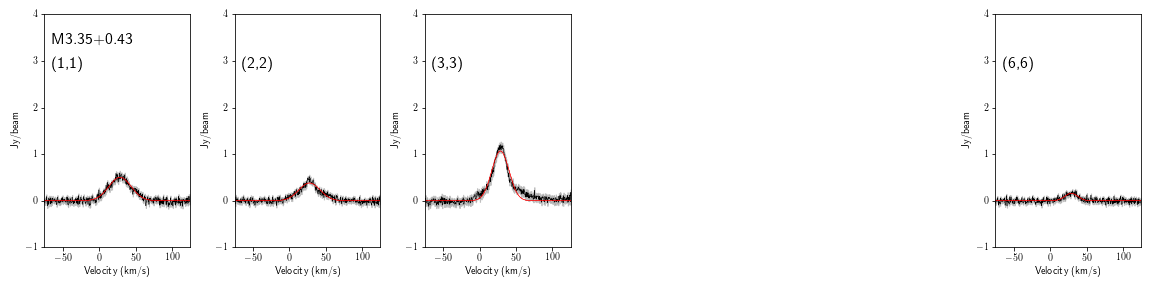}
    \caption{Source M3.34+0.43 showing (1,1)-(3,3) and (6,6) transitions.}
    \label{fig:fig3a}
\end{figure}

\begin{figure}[tbh]
    \centering
    \includegraphics[width=\textwidth]{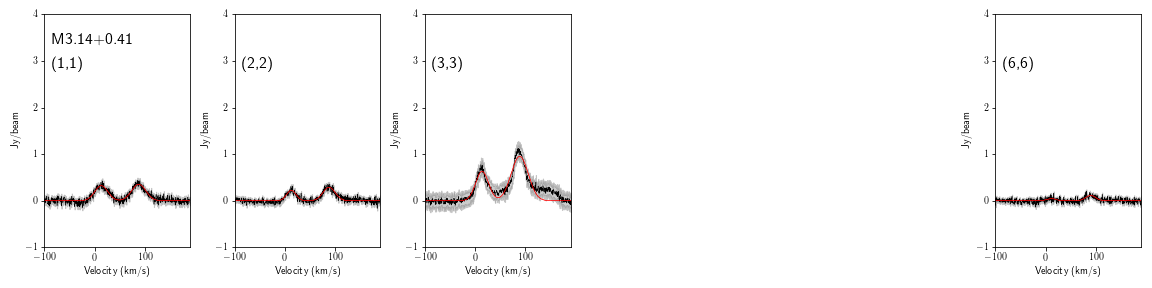}
    \caption{Source M3.14+0.41 showing (1,1)-(3,3) and (6,6) transitions.}
    \label{fig:fig3b}
\end{figure}

\begin{figure}[tbh]
    \centering
    \includegraphics[width=\textwidth]{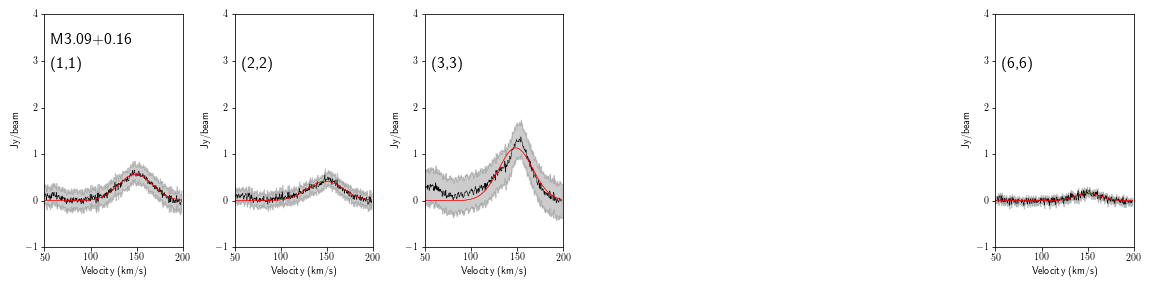}
    \caption{Source M3.09+0.16 showing (1,1)-(3,3) and (6,6) transitions.}
    \label{fig:fig3c}
\end{figure}

\begin{figure}[tbh]
    \centering
    \includegraphics[width=\textwidth]{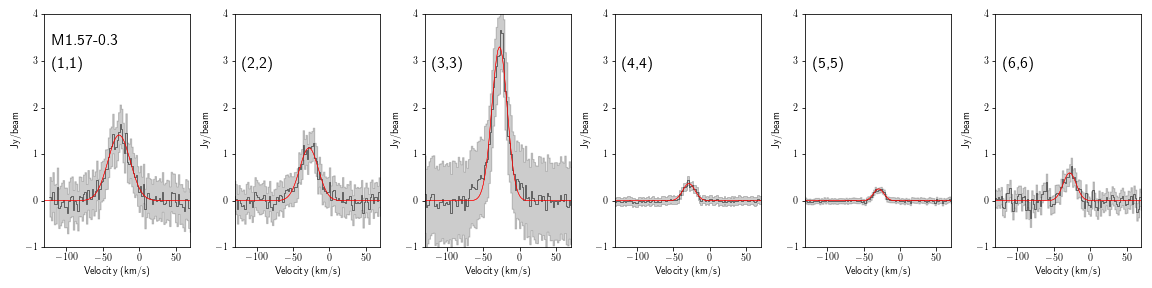}
    \caption{Source M1.57-0.3 showing (1,1)-(6,6) transitions.}
    \label{fig:fig3d}
\end{figure}

\begin{figure}[tbh]
    \centering
    \includegraphics[width=\textwidth]{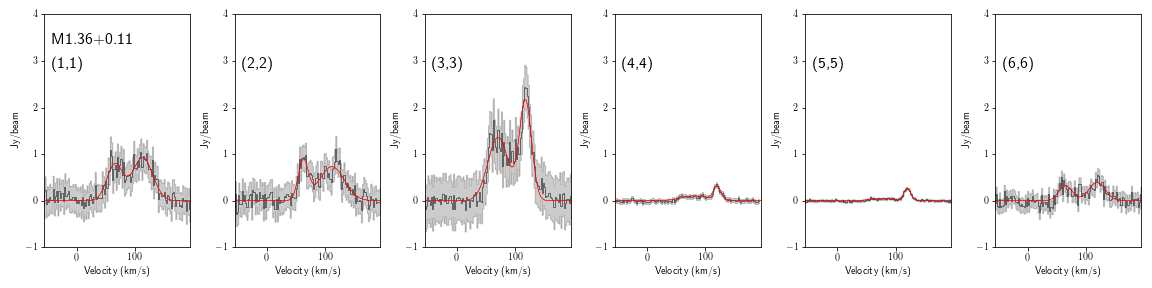}
    \caption{Source M1.36+0.11 showing (1,1)-(6,6) transitions.}
    \label{fig:fig3e}
\end{figure}

\begin{figure}[tbh]
    \centering
    \includegraphics[width=\textwidth]{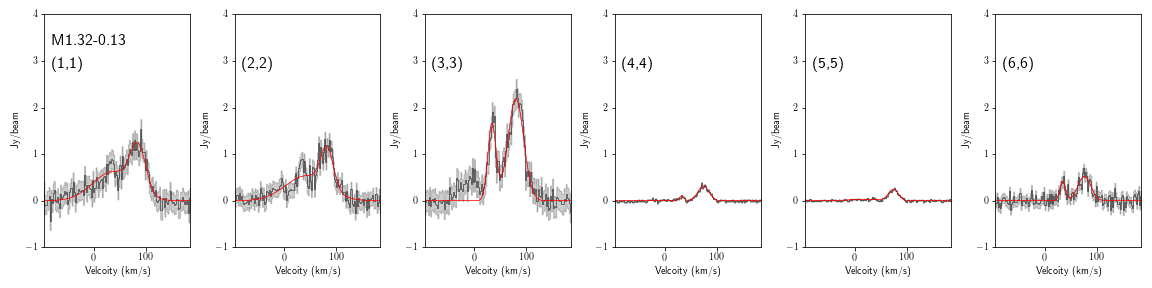}
    \caption{Source M1.32-0.13 showing (1,1)-(6,6) transitions.}
    \label{fig:fig3f}
\end{figure}

\begin{figure}[tbh]
    \centering
    \includegraphics[width=\textwidth]{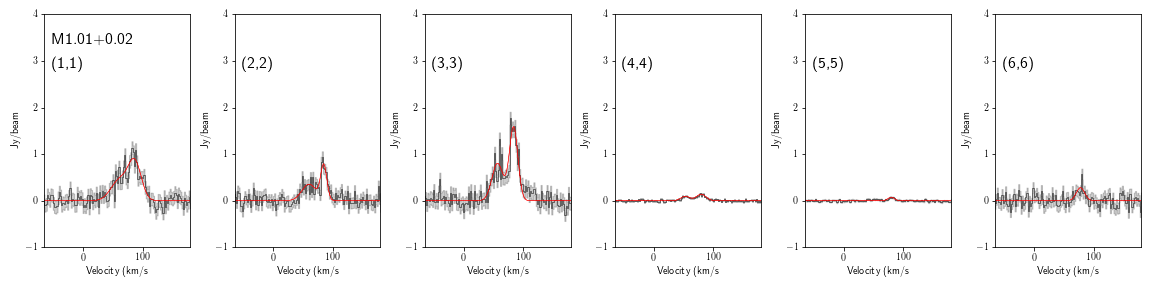}
    \caption{Source M1.01+0.02 showing (1,1)-(6,6) transitions.}
    \label{fig:fig3g}
\end{figure}

\begin{figure}[tbh]
    \centering
    \includegraphics[width=\textwidth]{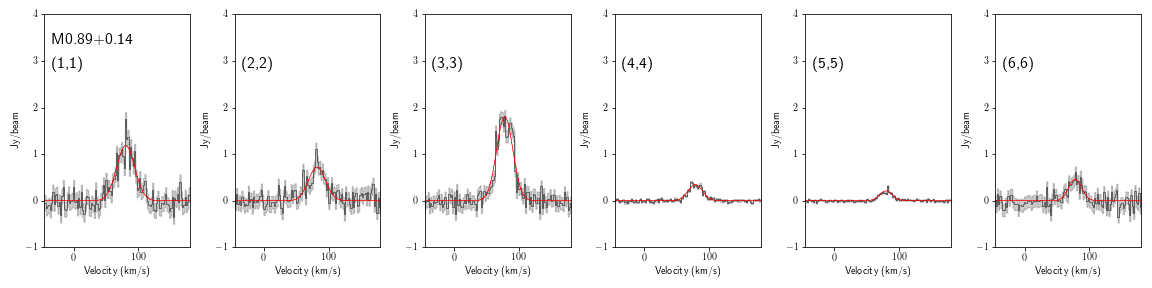}
    \caption{Source M0.89+0.14 showing (1,1)-(6,6) transitions.}
    \label{fig:fig3h}
\end{figure}

\begin{figure}[tbh]
    \centering
    \includegraphics[width=\textwidth]{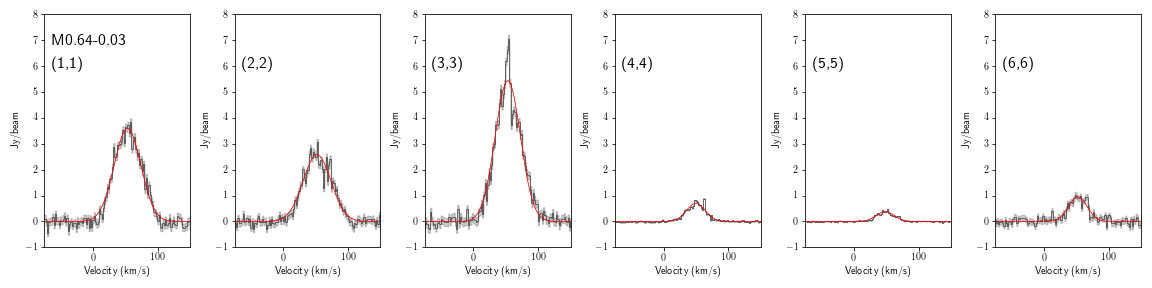}
    \caption{Source M0.64-0.03 showing (1,1)-(6,6) transitions.}
    \label{fig:fig3i}
\end{figure}

\begin{figure}[tbh]
    \centering
    \includegraphics[width=\textwidth]{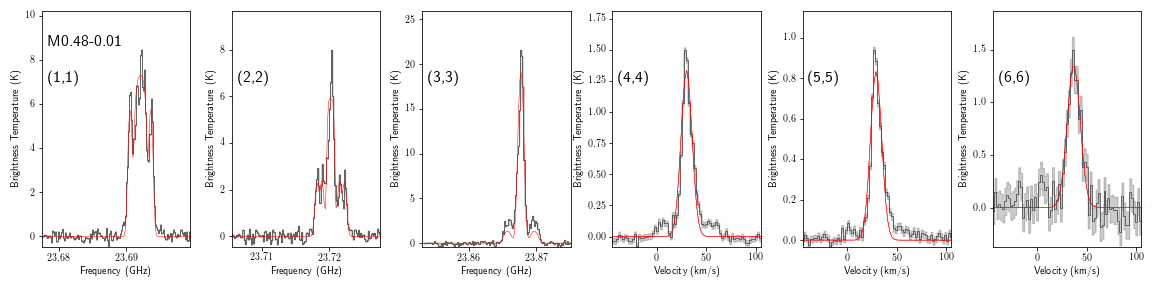}
    \caption{Source M0.48-0.01 showing (1,1)-(6,6) transitions.}
    \label{fig:fig3j}
\end{figure}

\begin{figure}[tbh]
    \centering
    \includegraphics[width=\textwidth]{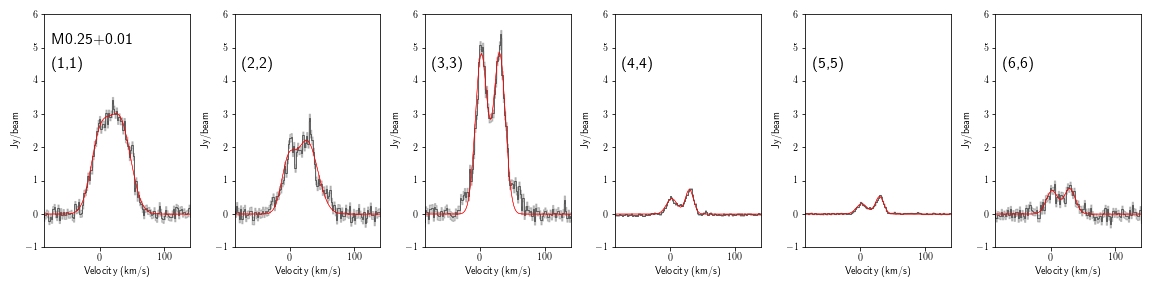}
    \caption{Source M0.25+0.01 showing (1,1)-(6,6) transitions.}
    \label{fig:fig3k}
\end{figure}

\begin{figure}[tbh]
    \centering
    \includegraphics[width=\textwidth]{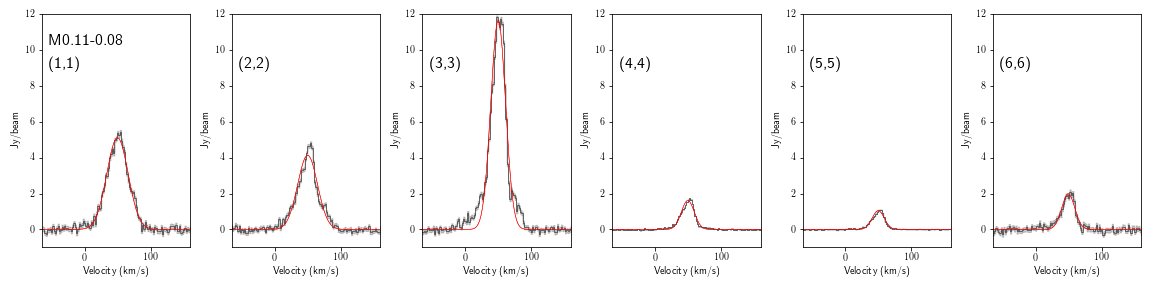}
    \caption{Source M0.11-0.08 showing (1,1)-(6,6) transitions.}
    \label{fig:fig3l}
\end{figure}

\begin{figure}[tbh]
    \centering
    \includegraphics[width=\textwidth]{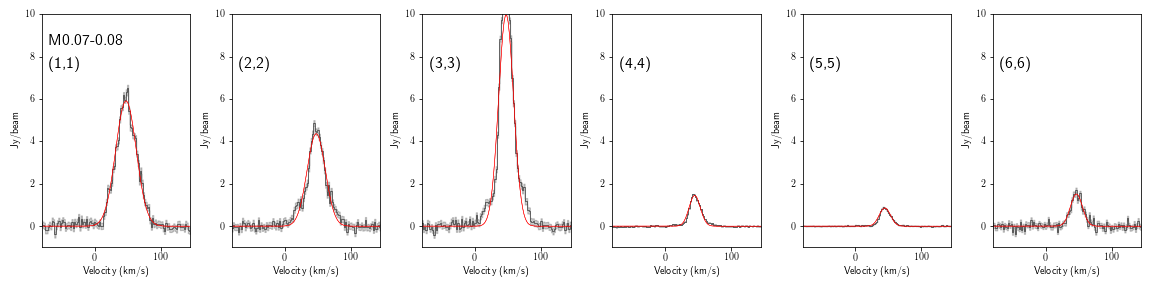}
    \caption{Source M0.07-0.08 showing (1,1)-(6,6) transitions.}
    \label{fig:fig3m}
\end{figure}

\begin{figure}[tbh]
    \centering
    \includegraphics[width=\textwidth]{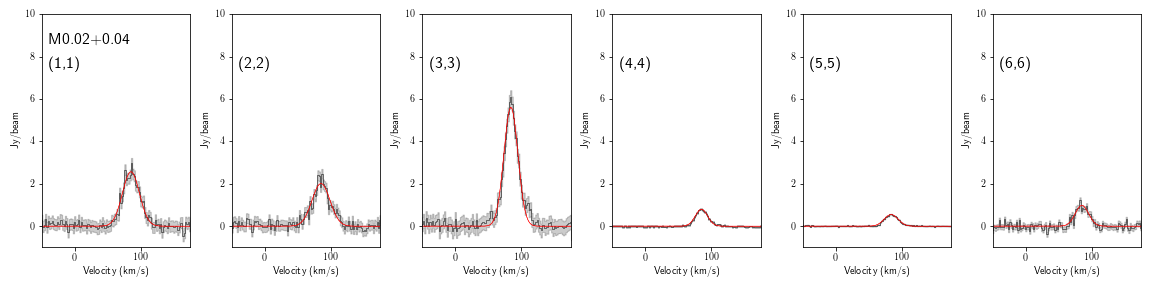}
    \caption{Source M0.02+0.04 showing (1,1)-(6,6) transitions.}
    \label{fig:fig3n}
\end{figure}

\begin{figure}[tbh]
    \centering
    \includegraphics[width=\textwidth]{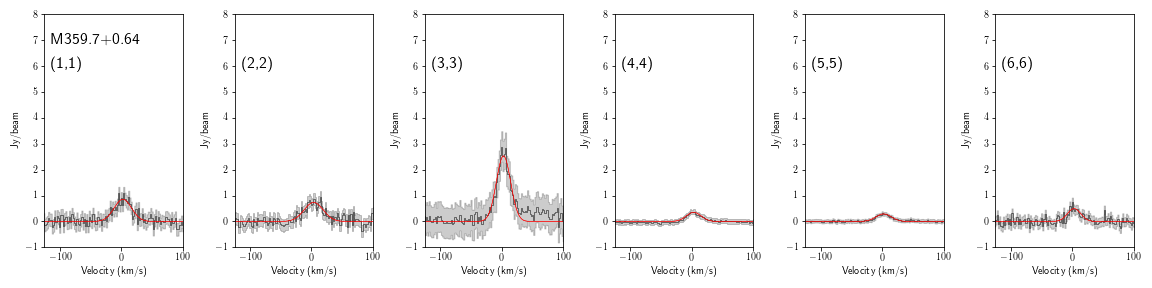}
    \caption{Source M359.7+0.64 showing (1,1)-(6,6) transitions.}
    \label{fig:fig3o}
\end{figure}

\begin{figure}[tbh]
    \centering
    \includegraphics[width=\textwidth]{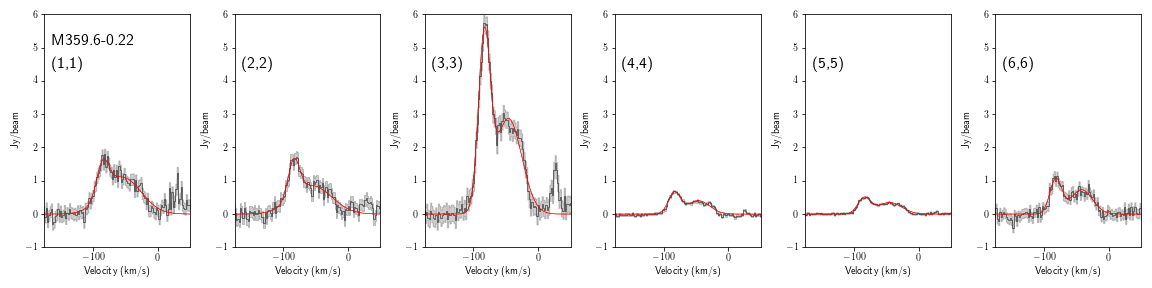}
    \caption{Source M359.6-0.22 showing (1,1)-(6,6) transitions.}
    \label{fig:fig3p}
\end{figure}

\clearpage 

\bibliography{candelaria.bib}
\bibliographystyle{aasjournal}

{}

\end{document}